\documentclass[aps,prx,twocolumn,superscriptaddress]{revtex4-2}
\usepackage{tabularx}
\usepackage{dcolumn}
\usepackage{graphics}% Include figure files
\usepackage{bm}% bold matha
\usepackage[dvips]{graphicx}
\usepackage[dvips]{rotating}
\usepackage[latin1]{inputenc}
\usepackage{dcolumn}
\usepackage{times}
\usepackage{tabularx}
\usepackage{amsmath}
\usepackage{amssymb}
\usepackage{float}
\usepackage{natbib}
\usepackage[dvipsnames]{xcolor}
\usepackage[normalem]{ulem}
\usepackage[colorlinks=true,linkcolor=blue,urlcolor=blue,citecolor=blue]{hyperref}

\newcommand\muB{\ensuremath{\mu_\textrm{B}}}
\newcommand\Vssd{\ensuremath{\hat{V}_\textrm{ss}}}
\newcommand\Vss{\ensuremath{\hat{V}_\textrm{SS}}}
\newcommand\bh[1]{\ensuremath{\hat{\textbf{#1}}}}
\begin{document}
\draft
\graphicspath{{./Figures/}}

\title{Spin--spin interaction and magnetic Feshbach resonances in collisions of high-spin atoms with closed-shell atoms}

\author{Matthew D. Frye}
\email{matthew.frye@fuw.edu.pl} 
\affiliation{Faculty of Physics, University of Warsaw, Pasteura 5, 02-093 Warsaw, Poland}
\author{Piotr S. {\.Z}uchowski}
\affiliation{Institute of Physics, Faculty of Physics, Astronomy and Informatics, Nicolaus Copernicus University, ul.\ Grudziadzka 5/7, 87-100
Torun, Poland}
\author{Micha{\l} Tomza}
\email{michal.tomza@fuw.edu.pl} 
\affiliation{Faculty of Physics, University of Warsaw, Pasteura 5, 02-093 Warsaw, Poland}

\date{\today}

\begin{abstract}

We investigate interactions and ultracold collisions of high-spin but spherical atoms with closed-shell atoms, focussing on the example of Cr with Yb. Such a combination has only one potential energy curve but gives rise to a substantial intra-atomic spin--spin interaction, which causes Feshbach resonances due to rotationally excited states. We find such resonances are guaranteed below 250 G and can reach widths of 10s of G in favorable circumstances. We study the effect of hyperfine structure, which can create additional wide resonances at low fields.
Finally, we consider isotopic substitution and show that for any realistic potential it is very likely at least one combination of commonly used isotopes will have resonances with experimentally favorable properties.
Thus, this is a promising system for both molecule formation and studying quantum mixtures including dipolar species.

\end{abstract}

\maketitle

\section{Introduction}

There is currently great interest in studying dipolar physics with ultracold systems. The long-range anisotropic dipole-dipole interaction opens up many avenues of new research. These include novel many-body physics such as quantum droplets and supersolidity \cite{Bottcher:2020, Chomaz:dipolar:2022, Bland:vortices:2023}, spin dynamics in lattices \cite{Li:tunable:2023}, and applications in quantum computing and simulation \cite{Ni:2018, Hughes:2020, Kaufman:2021, Altman:2021, Bao:dipolar:2023}. The physical systems used to realize such dipolar interactions include highly magnetic atoms -- such as Cr, Dy, and Er -- and heteronuclear molecules with permanent electric dipole moments.

Magnetic Feshbach resonances \cite{Chin:RMP:2010} often play an important role in such studies. They can be used to tune the strength of a contact interaction, allowing the dipolar nature of the interaction to be brought out \cite{Tanzi:2019, Hertkorn:2021, Patscheider:2022}. They also form the basis of magnetoassociation \cite{Kohler:RMP:2006}. This is where the magnetic field is tuned across a resonance, such that atom pairs coherently follow the resonant state as they cross the threshold and become a weakly bound state; this forms a crucial step in forming ultracold molecules from atoms, combined with coherent optical transfer to the ground state \cite{Vitanov:2017}. Understanding the properties of such Feshbach resonances is thus crucial for many aspects of this research.

One specific area of interest is in ground-state molecules with both electric and magnetic dipole moments. These give additional possibilities for control beyond closed-shell molecules due to the additional spin \cite{Micheli:2006, Baranov:2012, Karra:2016, Safronova:2018}.
One avenue for forming such molecules is through magnetoassociation of alkali-metal with closed-shell atoms. The Feshbach resonances in these systems are due to the weak distance-dependence of the hyperfine coupling and are sparse in field and extremely narrow \cite{Zuchowski:RbSr:2010, Brue:LiYb:2012, Brue:AlkYb:2013, Yang:CsYb:2019}. They have been observed for Rb+Sr \cite{Barbe:RbSr:2018}, Li+Yb \cite{Green:LiYb-res:2020}, and Cs+Yb \cite{Franzen:CsYb-res:2022}, but have not yet been used for magnetoassociation \cite{Borkowski:2023}. Another route is through direct laser cooling of molecules, which was initially demonstrated for simple diatomic species such as CaF \cite{Truppe:MOT:2017, Anderegg:2018}, SrF \cite{Barry:2014, McCarron:2018}, and YO \cite{Ding:2020, Burau:2023}, but has recently been extended to polyatomics \cite{Augenbraun:2023}.

More recently, attention has extended towards the possibility of forming molecules with high-spin atoms \cite{Pavlovic:2010, Tomza:2013, Tomza:2014, Zaremba:2018, Kosicki:ErYb:2020, Frye:DyYb:2020, Ciamei:2022, Soave:DyK:2023, Finelli:LiCr:2024}. The larger spins give both richer internal structure and larger magnetic dipole moments. However, the complexity of the interatomic interactions can pose problems for interpreting experiments and understanding the system well enough for molecule formation and quantum few-body applications.

The purpose of the present paper is to investigate the interactions and ultracold scattering of high-spin spherical atoms with closed-shell atoms. We take Cr($^7$S)+Yb ($^1$S) as our example system.
The intent is to find a class of systems with the simplicity and understandable structure of alkali-metal + closed-shell systems but with much stronger resonances.
We perform electronic structure calculations and show that the spin--spin interaction provides a substantial coupling for this system. We then perform scattering calculations and find that the Feshbach resonances caused by this coupling have useful widths and are guaranteed to exist at low to moderate magnetic fields. We also find particularly good tuning of the properties by isotopic substitution, so this system has significant insurance against unfavorable scattering lengths in any one isotopic combination. These results show Cr+Yb to be a promising system both for controlling interaction strengths and for molecule formation and show promise for a wider class of similar systems.

The structure of the paper is as follows. Section~\ref{sec:theory} introduces the theoretical and computational methods we employ. Section~\ref{sec:results} presents the main results and their robustness. Section~\ref{sec:conclusions} summarizes our findings and prospects.

\section{Theory}
\label{sec:theory}

The interaction of ground-state Cr and Yb produces only a single electronic interaction potential. This contrasts with the scattering of pairs of alkali-metal atoms, where there are two interaction potentials and the scattering dynamics is dominated by the coupling due to their splitting. Our present case is structurally simpler and as a result lacks this mechanism. Couplings in this system must therefore come from other interactions.

\subsection{Spin--spin interaction}

The spin--spin interaction arises between the magnetic dipoles of two or more unpaired electrons. There are a number of ways of representing this coupling, which are useful in different circumstances. The first we consider is suitable for the direct interaction of any two electrons, \cite{Brown:2003}
\begin{equation}
\Vssd=-g_s^2 \mu_\textrm{B}^2 \frac{\mu_0}{4\pi} \frac{3}{r^3} T^2(\hat{s}_1,\hat{s}_2)\cdot T^2(\bh{r},\bh{r}),
\label{eq:s1s2}
\end{equation}
where $\hat{s}_1$ and $\hat{s}_2$ represent the two interacting electron spins, and $r$ and $\bh{r}$ are the magnitude and direction of the vector between them. $g_s$ is the electron g-factor, $\mu_\textrm{B}$ is the Bohr magneton, and $\mu_0$ is the permeability of free space. $T^2(\cdot,\cdot)$ is a rank 2 tensor product. In order to consider the effects on atomic scattering, we must have an operator that acts on atomic coordinates rather than electronic ones. In systems where the two interacting spins are centered on different collision partners, it may be appropriate to simply substitute the atomic separation $\bf{R}$ for the electron separation $\bf{r}$; this form of the operator is commonly used in scattering of pairs of $^2$S atoms, including hydrogen and alkali-metal atoms \cite{Stoof:1988, Mies:1996, Hutson:Cs2:2008}, and is thus familiar in the theory of ultracold atoms.

When the spins involved in the coupling are centered on the same collision partner, or Eq.\ \eqref{eq:s1s2} is otherwise unsuitable, a molecular form of the operator is often appropriate. This can be written in terms of the coupled total spin $\hat{S}$ of all electrons involved in the coupling as \cite{Brown:2003}
\begin{equation}
\Vss=2\lambda(R) T^2(\hat{S},\hat{S})\cdot T^2(\bh{R},\bh{R}), \label{eq:SS}
\end{equation}
where the parameter $\lambda(R)$ generally needs to be obtained numerically or fitted to experiment \footnote{The strength of this coupling is sometimes described by the so-called zero-field splitting parameter $D$, particularly in the electron paramagnetic resonance community; this is related by $\lambda=D/2$.}.
There is also a well-known contribution to $\lambda(R)$ from the spin-orbit in second order; both contributions can be obtained from electronic structure calculations \cite{Vahtras:2002, Neese:ZFS:2007}. In a free atom, this coupling is zero by symmetry, but if another atom approaches and the orbitals of two electrons are distorted differently, it takes on a non-zero value. Due to the close contact between the electrons on the same atom, the interaction can become relatively large in this case.
This is the form of the operator we use throughout this paper; it
is commonly used in molecular spectroscopy \cite{Brown:2003}, but it has not been used in ultracold atomic scattering.

\begin{figure}[tbp]
\includegraphics[width=0.99\linewidth]{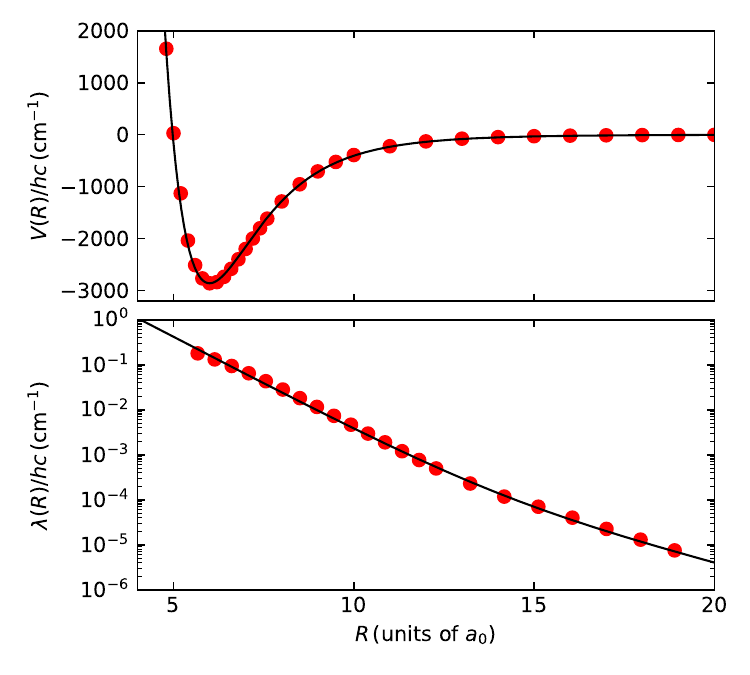}
\caption{Electronic interaction potential $V_\textrm{elec}(R)$ of Ref.\ \cite{Tomza:2013} (top) and spin--spin coupling coefficient $\lambda(R)$ (bottom) for Cr+Yb. We show both electronic structure (red points) and fitted functional forms (black). \label{fig:potential}}
\end{figure}

The interaction potential $V_\textrm{elec}(R)$ for Cr+Yb has been calculated previously in Ref.~\cite{Tomza:2013}. It has a depth of about $D_\textrm{e}=2860$ cm$^{-1}$, the leading long-range term $-C_6R^{-6}$ has dispersion coefficient $C_6=1195$ $E_\textrm{h}\,a_0^6$, and it supports 75 vibrational bound states. We calculate the spin--spin coupling coefficient $\lambda(R)$, taking into account all unpaired electrons, using the time-dependent density function theory framework developed by Neese \cite{Neese:ZFS:2007} and   implemented in the ADF package \cite{TeVelde:ADF:2001, ADF}. For these calculations, we use the ZORA scalar relativistic approach \cite{vanLenthe:1994} and QZ4P basis set~\cite{LentheJCC03} with the PBE functional \cite{Perdew:1996}, which works well for molecular properties.

The potential and the calculated coupling are shown in Fig.~\ref{fig:potential}. For use in scattering, we fit $V_\textrm{elec}(R)$ to a Morse-Long-Range functional form \cite{LeRoy:2007} and the spin--spin coupling to a double exponential decay; details are given in Supplemental Material \cite{supp_mat}.
This gives a value of $\lambda=0.41$ cm$^{-1}$ at the inner turning point.
%, which can be compared to 0.23 cm$^{-1}$ for CrH at its equilibrium \cite{Brown:2003, Corkery:CrH:1991}.
These fitted functions are also shown in Fig.\ \ref{fig:potential}.

\subsection{Scattering}

The Hamiltonian for scattering of Cr and Yb, each in their ground electronic state, is written as
\begin{equation}
\hat{H}=-\frac{\hbar^2}{2\mu}\frac{d^2}{dR^2}+\frac{\hbar^2}{2\mu}\frac{\hat{L}^2}{R^2}+\hat{h}_\textrm{Cr}+V_\textrm{elec}(R)+\Vss(R).
\end{equation}
Here, $\mu$ is the collisional reduced mass, and $\hat{L}^2$ is the operator for rotation of the two atoms around each other, with the associated quantum number $L$ and projection $M_L$; we include only $L=0$ and $2$ functions here as the coupling is perturbative. Unless otherwise stated we use $^{52}\textrm{Cr}$ and $^{174}\textrm{Yb}$, so neither atom has nuclear spin. The Yb atom is thus structureless; the Cr atom has spin $S=3$, so has Zeeman interaction $\hat{h}_\textrm{Cr}=\muB B g_s \hat{S}_z$, where $B$ is the magnetic field defining the $z$ axis and $\hat{S}_z$ is the projection operator for $S$ along that axis, with quantum number $M_S$. Note that $S$ and $M_S$ are also the total spin and projection of the system. We use coupled-channels calculations for both scattering and bound states. The coupled equations are constructed and solved \cite{Alexander:1987, MG:symplectic:1995, CS4} using \textsc{molscat} \cite{molscat:2019,mbf-github:2022} for scattering calculations and \textsc{bound} and \textsc{field} \cite{bound+field:2019, mbf-github:2022} for bound-state calculations. Further details of the calculations are given in Supplemental Material \cite{supp_mat}.

The key features of the scattering that we are interested in are magnetic Feshbach resonances. These occur when a bound state crosses a threshold as a function of the magnetic field, and there is a suitable coupling between the incoming state and the bound state.
Although there is only a single interaction potential for this system, bound states can cross a threshold if they have different $M_S$ and therefore different Zeeman effects.
The coupling here is provided by the spin--spin interaction,
which in first order can change the partial wave by $\Delta L=0,\pm 2$, but has no matrix elements between $L=L'=0$ functions. It also changes $\Delta M_S=0,\pm 1,\pm 2$, with a compensating change in $M_L$. This suggests that for s-wave scattering in the lowest initial state -- $L=0$, $M_S=-3$ -- the candidate states to cause Feshbach resonances are those with $L=2$ and $M_S=-2$ or $-1$.

The signature of a magnetic Feshbach resonance is a pole in the scattering length, $a(B)=a_\textrm{bg}\left[1-\Delta/(B-B_\textrm{res})\right]$ \cite{Moerdijk:1995}. The resonance is described by three parameters: $B_\textrm{res}$ is the resonance position, $\Delta$ is its field width, and $a_\textrm{bg}$ is the background scattering length far from resonance. With these definitions, the width $\Delta$ can be artificially large or small when $a_\textrm{bg}$ is particularly small or large, respectively. We instead follow Ref.\ \cite{Yang:CsYb:2019} and use a normalised width $\bar{\Delta}=a_\textrm{bg}\Delta/\bar{a}$,
where $\bar{a}$ is the mean scattering length \cite{Gribakin:1993}, which is $55\,a_0$ here. This is a better measure than $\Delta$ itself, both of the coupling strength and of the use of the resonance for tuning scattering lengths or magnetoassociation. We use the algorithms of Frye and Hutson \cite{Frye:resonance:2017,Frye:quasibound:2020} to locate and characterize the resonances in our scattering calculations.

\begin{figure}[tbp]
\includegraphics[width=0.99\linewidth]{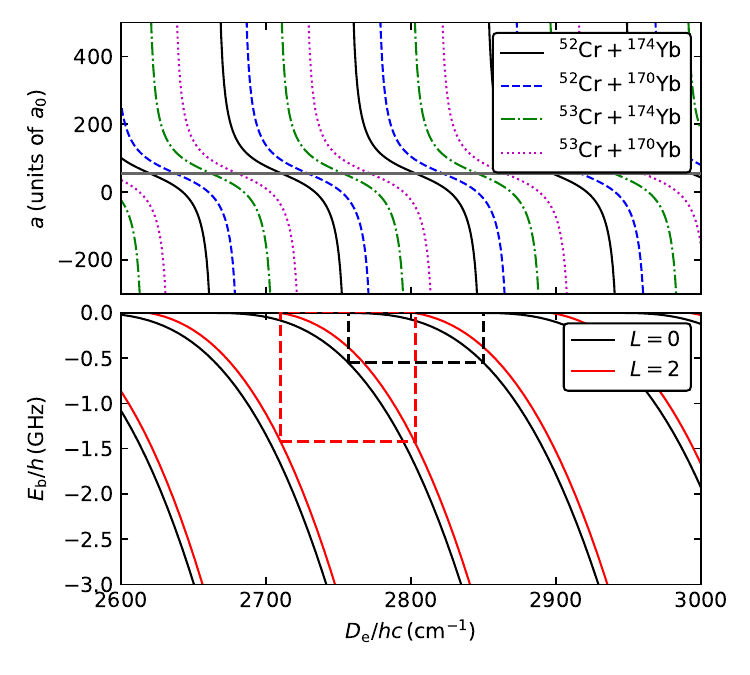}
\caption{Top: the s-wave scattering length for four isotopic combinations. Bottom: near-threshold bound states for $^{52}\textrm{Cr}+{^{174}\textrm{Yb}}$ for $L=0$ (black) and 2 (red). Both are shown as the depth of the potential, $D_\textrm{e}$, is varied. The horizontal grey line indicates $\bar{a}$. The dashed boxes show the depth of the top bin for each $L$ and the variation in $D_\textrm{e}$ that corresponds to a cycle of scattering length.
\label{fig:single_channel}}
\end{figure}

\section{Results}
\label{sec:results}

\subsection{Single channel}

As a preliminary step, we consider the system without internal structure or spin--spin coupling.
This is a useful first-order picture because all the spin-dependent terms in the Hamiltonian are relatively weak and expected to act perturbatively. Figure~\ref{fig:single_channel} shows the s-wave scattering length $a$ and near-threshold bound states as a function of $D_\textrm{e}$.
The top panel shows the usual behavior of the scattering length: it is cyclic, running between $\pm\infty$ but most likely to be around $\bar{a}$ \cite{Gribakin:1993}.

The lower panel of Fig.\ \ref{fig:single_channel} shows the corresponding near-threshold bound states. As expected, these have the same periodicity as the scattering length, with an s-wave bound state crossing threshold at poles in $a$. The d-wave bound states also have the same periodicity but offset by half a cycle such that one crosses the threshold when $a=\bar{a}$ \cite{Gao:2000}. There is always a single bound state in a certain ``bin'' of energy below the threshold, which is 540 MHz for $L=0$ and 1400 MHz for $L=2$ \cite{Gao:2000}.
The uncertainty of $D_\textrm{e}$ was estimated to be 10\% \cite{Tomza:2013}, which covers several cycles of $a$. We, therefore, cannot predict the true value of $a$ nor position of the bound states, but we can use this cyclic dependence to explore the behavior over one complete cycle; because this system has only a single potential, this procedure covers the entire range of possible behaviors of this system and so can stand in for any possible true potential.
Therefore, in the remainder of the paper, we scale $D_\textrm{e}$ from 2757 to 2850 cm$^{-1}$ to cover a single full cycle of $a$ from $-\infty$ to $+\infty$; this range is indicated by the black box in Fig.\ \ref{fig:single_channel} and for convenience we map this to a scaling from 0 to 1.
We also plot $a$ for 3 other isotopic combinations, showing that this cycle can be shifted across most of a period by isotope substitution; we will return to this idea later.

\begin{figure}[tbp]
\includegraphics[width=0.99\linewidth]{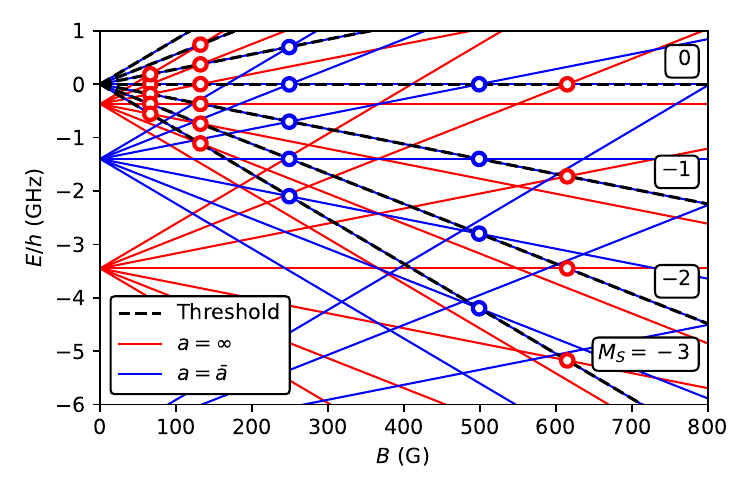}
\caption{Thresholds and near-threshold bound states with $L=2$ for $^{52}\textrm{Cr}+{^{174}\textrm{Yb}}$, for two scalings of the potential giving $a=\infty$ (red) and $a=\bar{a}$ (blue); the least-bound state for $a=\bar{a}$ is at zero binding energy (as shown in Fig.\ \ref{fig:single_channel}) so the corresponding lines lie directly below those for the thresholds. Crossings that can cause Feshbach resonances, with selection rules as described in the text, are shown as dots. \label{fig:crossing}}
\end{figure}

We now look to the patterns of bound states that can cause Feshbach resonances. As discussed above, the states of most interest are d-wave ($L=2$), and these are shown in Fig.\ \ref{fig:crossing} as a function of magnetic field, for two scattering lengths $a=\infty$ and $a=\bar{a}$.
In the absence of spin--spin coupling, $M_S$ is a good quantum number so the bound states lie parallel to the thresholds which support them.
Resonances can occur where these states cross thresholds and the selection rules for the spin--spin coupling are satisfied; these locations are also marked.
For the case of $a=\bar{a}$ (blue lines), there is a d-wave state right at the threshold and the next at the bottom of the top bin, see Fig.\ \ref{fig:single_channel}.
The true least-bound state must lie between these two states, so the resonance caused by the lower state represents the maximum field at which the first resonance will appear; this field
is $B_\textrm{res}=E_\textrm{b}/(\mu_\textrm{B}g_s\Delta m_s)=250$ G for $\Delta m_s=+2$ and 500 G for $\Delta m_s=+1$.

\begin{figure}[tbp]
\includegraphics[width=0.99\linewidth]{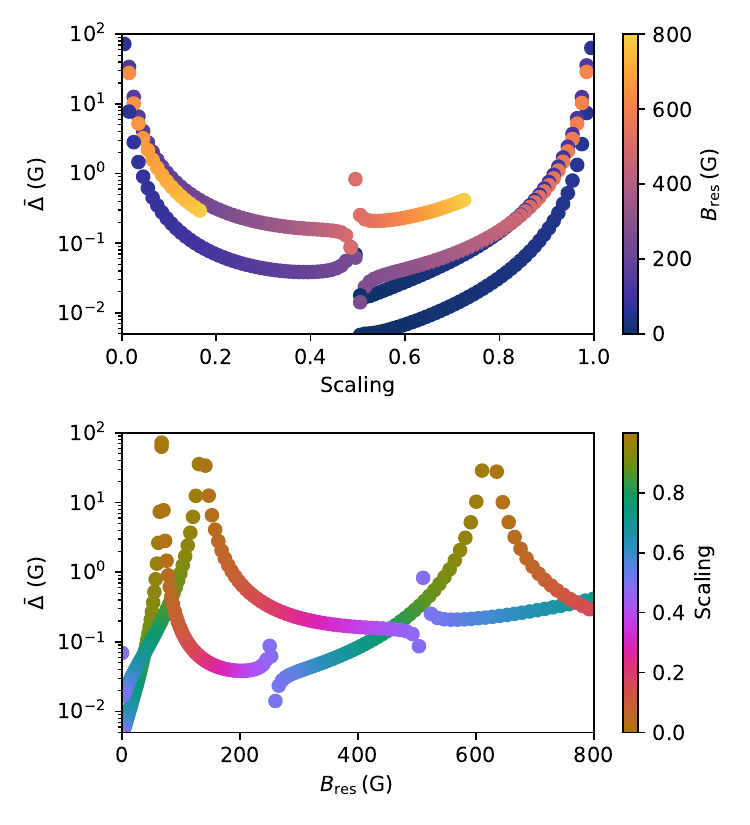}
\caption{Normalised widths $\bar{\Delta}$ of the characterised resonances for $^{52}\textrm{Cr}+{^{174}\textrm{Yb}}$ for 100 potentials.
The scaling runs from $D_\textrm{e} = 2757$ to 2850 cm$^{-1}$ and covers the full cycle of scattering length from $-\infty$ to $+\infty$ as marked by the black box in Fig.\ \ref{fig:single_channel} and discussed in text.
Both panels present the same results: the upper panel shows $\bar{\Delta}$ as a function of potential scaling, with $B_\textrm{res}$ encoded in the color of the points; and the lower panel shows $\bar{\Delta}$ as a function of $B_\textrm{res}$, with the scaling encoded in the color of the points. \label{fig:parameters}}
\end{figure}

\subsection{Resonance results}

We locate and characterize each resonance at the lowest $M_S=-3$ threshold while varying the potential.
We use 100 different potentials, with $D_\textrm{e}$ equally spaced across the cycle of scattering length as discussed above and marked in Fig.\ \ref{fig:single_channel}, and we characterize all resonances up to 800 G for each potential.
These results are presented in Fig.\ \ref{fig:parameters}; full tabulations of resonance parameters are contained in Supplemental Material \cite{supp_mat}.
We can see that resonances are quite likely to occur at very low fields, as suggested from Fig.\ \ref{fig:crossing}, but they are correspondingly narrower.
This occurs because the bound states that can cause resonances at such low fields are extremely weakly bound in their own channels, and so extend to very long range and have less probability density in the short-range region where the coupling is significant \cite{Brue:AlkYb:2013}.
Conversely, the widths have peaks at 65 and 130 G, which reach over 10 G width, due to the large s-wave scattering length around scalings of 1 or 0, which enhances the wavefunction at short range and thus the width \cite{Brue:AlkYb:2013}. Even away from these peaks, there are resonances with usefully large widths ($\bar{\Delta}>0.1$ G) for most scalings. The resonances for $\Delta M_S=+1$ are typically wider than the corresponding ones for $\Delta M_S=+2$ -- due to angular momentum factors and a smaller relative magnetic moment -- but are also at higher field; the 65 and 130 G peaks thus correspond to $\Delta M_S=+2$ and $+1$ resonances respectively, as shown by red points in Fig.\ \ref{fig:crossing}. The oscillation in width around scaling of 0.5 (250 and 500 G) is related to a d-wave shape resonance.

The characteristics of widths shown here are determined by only the (accurately calculated) $C_6$ coefficient, together with known atomic properties. The patterns of widths are, therefore, very robust predictions. It is unknown \emph{a priori} whether a particular system does indeed have a resonance in a particular field range -- it depends on the short-range potential, or equivalently on the scattering length -- but we can say with certainty that \emph{if} a resonance does exist at a specific field it will have properties as predicted here. For example, $\Delta m_f=+1$ resonances near 130 G will always occur for potentials with large $a$, so they will always have greatly enhanced widths. The main source of uncertainty here is our calculated $\lambda(R)$, which scales the widths in a simple (perturbative) way and does not change the overall pattern.
In fact, very similar patterns will occur for any system of a high-spin spherical atom colliding with a closed-shell atom.
There will be a different scaling of the magnetic fields (through the binding energies, as shown in Fig.\ \ref{fig:crossing}) and widths due to different $C_6$, $\mu$, and $\lambda(R)$ for different systems \cite{Gao:2000, Brue:AlkYb:2013}, but the overall pattern of the widths is universal. 

\subsection{Effects of hyperfine coupling}

We now turn to the effects of hyperfine interactions on the scattering \footnote{Nuclear spin on the Yb atom may cause Feshbach resonances similar to those in alkali-metal plus closed-chell atom collisions \cite{Brue:LiYb:2012, Barbe:RbSr:2018, Yang:CsYb:2019}, but these are expected to be very narrow, so we consider only nuclear spin on the Cr.}. $^{53}$Cr has a nuclear spin of $i=3/2$, so we add the Fermi-contact interaction $A_\textrm{HF} \hat{S}\cdot\hat{i}$ to the scattering Hamiltonian. This couples $S$ and $i$ to form $f=3/2$, $5/2$, $7/2$, and $9/2$; at zero field, the states are spread over $\sim500$ MHz \cite{Reinhardt:Cr_HF:1995} above the ground state $f=9/2$.
These levels are split by a magnetic field into states which are characterized by $m_f$ at low field, and the selection rule for \Vss\ becomes $\Delta m_f = 0,\pm1,\pm2$. As the field increases, the spins decouple, and states are again characterized by $M_S$ and $m_i$; in this regime, the nuclear spin becomes mostly a spectator, and the system behaves similarly to the $i=0$ case.

\begin{figure}[tbp]
\includegraphics[width=0.99\linewidth]{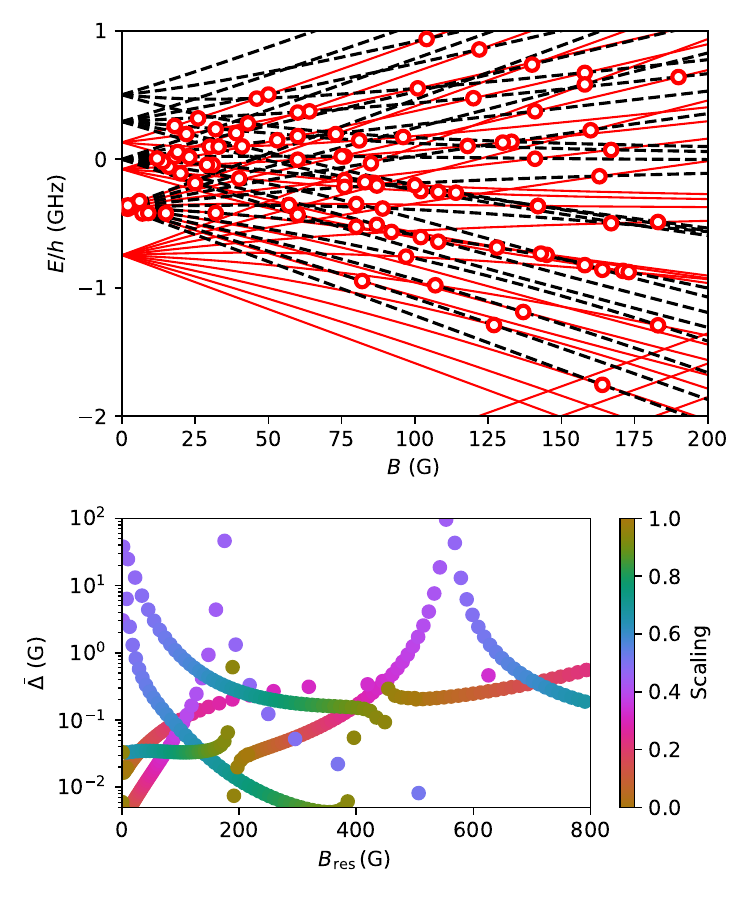}
\caption{Feshbach resonances for $^{53}\textrm{Cr}+{^{174}\textrm{Yb}}$. Top: Hyperfine crossing diagram, focussing on the low-field region; red lines and points show bound states and resonances for the case of $a=\infty$. Bottom: Normalised widths $\bar{\Delta}$ as a function of $B_\textrm{res}$, with the scaling encoded in the color of the points.
The scaling in this case covers scattering lengths from $a\approx\bar{a}$ through $\pm \infty$ and back to $\bar{a}$.
\label{fig:HF_combined}}
\end{figure}

Figure \ref{fig:HF_combined} shows results for $^{53}\textrm{Cr}+{^{174}\textrm{Yb}}$ including hyperfine coupling.
The top panel shows a crossing diagram equivalent to Fig.\ \ref{fig:crossing}, for $a=\infty$.
For states originating from the same $f=9/2$ hyperfine manifold, the crossings are shifted to a higher field for a given binding energy due to a reduction in the relative magnetic moment and the non-linear Zeeman effect. 
However, it is also possible to have resonances due to states from other hyperfine manifolds, and in this case, those with $f=7/2$ occur at very low fields.
The bottom panel of Fig.\ \ref{fig:HF_combined} shows the corresponding resonance widths in the lowest channel $(f,m_f)=(9/2,-9/2)$. Note that we use the same set of $D_e$ as previously, but due to the different reduced mass $\mu$ these now correspond to different scattering lengths; this set still covers a full cycle of scattering length, but instead of running from $+\infty$ to $-\infty$, it now runs approximately from $\bar{a}$ through $\pm\infty$ (close to scaling 0.5) and back to $\bar{a}$, see Fig.\ \ref{fig:single_channel}.
The most notable feature is the presence of low-field resonances with very large widths; these are caused by states with $f=7/2$, which cross the threshold of interest at low fields when $a$ is close to a pole and so have greatly enhanced widths as discussed above.
Whether this feature occurs in the real system is once again dependent on the (unknown) scattering length, but the possibility of this enhancement happening at low field relies on only the known atomic hyperfine structure and the accurately calculated $C_6$ coefficient and is not due to any quirk of our particular potential. As expected, at higher magnetic fields from about 400 G, the pattern of widths closely resembles that for $^{52}$Cr shown in Fig.\ \ref{fig:parameters}.

\subsection{Isotopic variation}

We have so far considered properties while varying the interaction potential; in the real system, the potential is fixed, but we can instead change the isotopes of either atom. This changes the reduced mass, which is conceptually very similar to scaling the interaction potential, but only discrete steps are available. This effect is shown in Fig.\ \ref{fig:single_channel} for combinations of the two most experimentally friendly isotopes of each species: $^{52}$Cr and $^{53}$Cr, and $^{170}$Yb and $^{174}$Yb. It can be seen that changing the Yb isotope shifts the scattering length by about 20\% of a cycle, and changing the Cr isotope shifts by 55\% of a cycle.

We repeat our calculations of resonance parameters for the four isotopic combinations above for each of our scaled potentials. Full tabulations of resonance parameters are again included in Supplemental Material \cite{supp_mat}. We find that for every potential, there is always at least one resonance below 300 G with $\bar{\Delta}>0.1$ G and a 2/5 chance of one with $\bar{\Delta}>1$ G. Even if we restrict that the background scattering length should be moderate and positive -- $0<a_\textrm{bg}<4\bar{a}$ -- then 95\% of potentials have a resonance below 300 G with $\bar{\Delta}>0.1$ G. Both species have further isotopes that could be used and would make these conclusions even more robust.

The interactions and resonances considered here will also occur in a range of similar systems. Any closed-shell atom could be used in place of Yb, for example, alkaline-earth metals such as Sr, which is readily prepared at ultracold temperatures. In place of Cr, there are a number of high-spin spherical atoms that could be used, such as Mo or Mn if they can be prepared at ultracold temperatures. Perhaps the most interesting possibility is Eu, which has an $^8$S ground state and two stable isotopes, each with rich hyperfine structure, and which was Bose-Einstien condensed recently~\cite{MiyazawaPRL22}. Our preliminary calculations suggest that the spin--spin coupling is similar or somewhat stronger in Eu+Yb than in Cr+Yb so that this system may be fertile ground for future work.

\section{Conclusions}
\label{sec:conclusions}

We have investigated interactions and collisions for a novel combination of high-spin spherical and closed-shell atoms, taking Cr+Yb as our example system. The primary coupling mechanism is the intra-atomic spin--spin interaction, which we calculate to be a substantial fraction of wavenumber in the interaction region. This couples rotationally excited states to s-wave scattering and creates magnetic Feshbach resonances. We locate and characterize these resonances in our scattering calculations and interpret the resulting patterns of widths, using a representative sample of interaction potentials to cover the range of possible scattering lengths. These show that such resonances are generally of the order 0.1 G wide at moderate fields of a few 100 G, but can reach much larger widths when enhanced by large background scattering lengths. When hyperfine is present in the system, there is additional potential for extremely broad resonances at very low fields.

Selecting among isotopes allows some control of scattering lengths and, thus, resonance properties, and we show that this system is extremely robust to unlucky potentials when as few as two isotopes of each species are considered. The precise details of which resonance is favorable depend on the unknown scattering lengths, but our conclusions depend on only known atomic quantities, the accurately calculated $C_6$ coefficient, and our $\lambda(R)$. Whatever the true potential, a choice of isotopes allows us to select for resonance properties that are experimentally favorable.

These Feshbach resonances will have applications both in controlling atomic mixtures and in molecule formation. Once formed in weakly bound Feshbach states, molecules may be optically transferred to the ground state.
The resulting ground-state molecules will have large magnetic moments inherited directly from the high-spin atom and electric dipole moments -- CrYb is predicted to have 1.2 D \cite{Tomza:2013}. These will have applications in a wide range of quantum computing and simulation settings in addition to fundamental interest in the behaviors of molecules with two dipole moments.

\begin{acknowledgments}
We gratefully acknowledge the Foundation for Polish Science within the First Team programme cofinanced by the European Union under the European Regional Development Fund and the National Science Center, Poland (grant no.~2020/38/E/ST2/00564) for the financial support and Poland's high-performance computing infrastructure PLGrid (HPC Centers: ACK Cyfronet AGH) for providing computer facilities and support (computational grant no.~PLG/2021/015237). PS\.Z was supported by the National Science Centre, Poland (grant no.~2021/41/B/ST2/00681).
\end{acknowledgments}

\bibliography{CrYb_resonance}

\begin{thebibliography}{71}%
\makeatletter
\providecommand \@ifxundefined [1]{%
 \@ifx{#1\undefined}
}%
\providecommand \@ifnum [1]{%
 \ifnum #1\expandafter \@firstoftwo
 \else \expandafter \@secondoftwo
 \fi
}%
\providecommand \@ifx [1]{%
 \ifx #1\expandafter \@firstoftwo
 \else \expandafter \@secondoftwo
 \fi
}%
\providecommand \natexlab [1]{#1}%
\providecommand \enquote  [1]{``#1''}%
\providecommand \bibnamefont  [1]{#1}%
\providecommand \bibfnamefont [1]{#1}%
\providecommand \citenamefont [1]{#1}%
\providecommand \href@noop [0]{\@secondoftwo}%
\providecommand \href [0]{\begingroup \@sanitize@url \@href}%
\providecommand \@href[1]{\@@startlink{#1}\@@href}%
\providecommand \@@href[1]{\endgroup#1\@@endlink}%
\providecommand \@sanitize@url [0]{\catcode `\\12\catcode `\$12\catcode
  `\&12\catcode `\#12\catcode `\^12\catcode `\_12\catcode `\%12\relax}%
\providecommand \@@startlink[1]{}%
\providecommand \@@endlink[0]{}%
\providecommand \url  [0]{\begingroup\@sanitize@url \@url }%
\providecommand \@url [1]{\endgroup\@href {#1}{\urlprefix }}%
\providecommand \urlprefix  [0]{URL }%
\providecommand \Eprint [0]{\href }%
\providecommand \doibase [0]{https://doi.org/}%
\providecommand \selectlanguage [0]{\@gobble}%
\providecommand \bibinfo  [0]{\@secondoftwo}%
\providecommand \bibfield  [0]{\@secondoftwo}%
\providecommand \translation [1]{[#1]}%
\providecommand \BibitemOpen [0]{}%
\providecommand \bibitemStop [0]{}%
\providecommand \bibitemNoStop [0]{.\EOS\space}%
\providecommand \EOS [0]{\spacefactor3000\relax}%
\providecommand \BibitemShut  [1]{\csname bibitem#1\endcsname}%
\let\auto@bib@innerbib\@empty
%</preamble>
\bibitem [{\citenamefont {B\"ottcher}\ \emph {et~al.}(2020)\citenamefont
  {B\"ottcher}, \citenamefont {Schmidt}, \citenamefont {Hertkorn},
  \citenamefont {Ng}, \citenamefont {Graham}, \citenamefont {Guo},
  \citenamefont {Langen},\ and\ \citenamefont {Pfau}}]{Bottcher:2020}%
  \BibitemOpen
  \bibfield  {author} {\bibinfo {author} {\bibfnamefont {F.}~\bibnamefont
  {B\"ottcher}}, \bibinfo {author} {\bibfnamefont {J.-N.}\ \bibnamefont
  {Schmidt}}, \bibinfo {author} {\bibfnamefont {J.}~\bibnamefont {Hertkorn}},
  \bibinfo {author} {\bibfnamefont {K.~S.~H.}\ \bibnamefont {Ng}}, \bibinfo
  {author} {\bibfnamefont {S.~D.}\ \bibnamefont {Graham}}, \bibinfo {author}
  {\bibfnamefont {M.}~\bibnamefont {Guo}}, \bibinfo {author} {\bibfnamefont
  {T.}~\bibnamefont {Langen}},\ and\ \bibinfo {author} {\bibfnamefont
  {T.}~\bibnamefont {Pfau}},\ }\bibfield  {title} {\bibinfo {title} {New states
  of matter with fine-tuned interactions: quantum droplets and dipolar
  supersolids},\ }\href {https://doi.org/10.1088/1361-6633/abc9ab} {\bibfield
  {journal} {\bibinfo  {journal} {Rep. Prog. Phys.}\ }\textbf {\bibinfo
  {volume} {84}},\ \bibinfo {pages} {012403} (\bibinfo {year}
  {2020})}\BibitemShut {NoStop}%
\bibitem [{\citenamefont {Chomaz}\ \emph {et~al.}(2022)\citenamefont {Chomaz},
  \citenamefont {Ferrier-Barbut}, \citenamefont {Ferlaino}, \citenamefont
  {Laburthe-Tolra}, \citenamefont {Lev},\ and\ \citenamefont
  {Pfau}}]{Chomaz:dipolar:2022}%
  \BibitemOpen
  \bibfield  {author} {\bibinfo {author} {\bibfnamefont {L.}~\bibnamefont
  {Chomaz}}, \bibinfo {author} {\bibfnamefont {I.}~\bibnamefont
  {Ferrier-Barbut}}, \bibinfo {author} {\bibfnamefont {F.}~\bibnamefont
  {Ferlaino}}, \bibinfo {author} {\bibfnamefont {B.}~\bibnamefont
  {Laburthe-Tolra}}, \bibinfo {author} {\bibfnamefont {B.~L.}\ \bibnamefont
  {Lev}},\ and\ \bibinfo {author} {\bibfnamefont {T.}~\bibnamefont {Pfau}},\
  }\bibfield  {title} {\bibinfo {title} {Dipolar physics: a review of
  experiments with magnetic quantum gases},\ }\href
  {https://doi.org/10.1088/1361-6633/aca814} {\bibfield  {journal} {\bibinfo
  {journal} {Rep. Prog. Phys.}\ }\textbf {\bibinfo {volume} {86}},\ \bibinfo
  {pages} {026401} (\bibinfo {year} {2022})}\BibitemShut {NoStop}%
\bibitem [{\citenamefont {Bland}\ \emph {et~al.}(2023)\citenamefont {Bland},
  \citenamefont {Lamporesi}, \citenamefont {Mark},\ and\ \citenamefont
  {Ferlaino}}]{Bland:vortices:2023}%
  \BibitemOpen
  \bibfield  {author} {\bibinfo {author} {\bibfnamefont {T.}~\bibnamefont
  {Bland}}, \bibinfo {author} {\bibfnamefont {G.}~\bibnamefont {Lamporesi}},
  \bibinfo {author} {\bibfnamefont {M.~J.}\ \bibnamefont {Mark}},\ and\
  \bibinfo {author} {\bibfnamefont {F.}~\bibnamefont {Ferlaino}},\ }\bibfield
  {title} {\bibinfo {title} {Vortices in dipolar {B}ose-{E}instein
  condensates},\ }\href {https://doi.org/10.5802/crphys.160} {\bibfield
  {journal} {\bibinfo  {journal} {C. R. Physique}\ }\textbf {\bibinfo {volume}
  {24}} (\bibinfo {year} {2023})}\BibitemShut {NoStop}%
\bibitem [{\citenamefont {Li}\ \emph {et~al.}(2023)\citenamefont {Li},
  \citenamefont {Matsuda}, \citenamefont {Miller}, \citenamefont {Carroll},
  \citenamefont {Tobias}, \citenamefont {Higgins},\ and\ \citenamefont
  {Ye}}]{Li:tunable:2023}%
  \BibitemOpen
  \bibfield  {author} {\bibinfo {author} {\bibfnamefont {J.-R.}\ \bibnamefont
  {Li}}, \bibinfo {author} {\bibfnamefont {K.}~\bibnamefont {Matsuda}},
  \bibinfo {author} {\bibfnamefont {C.}~\bibnamefont {Miller}}, \bibinfo
  {author} {\bibfnamefont {A.~N.}\ \bibnamefont {Carroll}}, \bibinfo {author}
  {\bibfnamefont {W.~G.}\ \bibnamefont {Tobias}}, \bibinfo {author}
  {\bibfnamefont {J.~S.}\ \bibnamefont {Higgins}},\ and\ \bibinfo {author}
  {\bibfnamefont {J.}~\bibnamefont {Ye}},\ }\bibfield  {title} {\bibinfo
  {title} {Tunable itinerant spin dynamics with polar molecules},\ }\href
  {https://doi.org/10.1038/s41586-022-05479-2} {\bibfield  {journal} {\bibinfo
  {journal} {Nature}\ }\textbf {\bibinfo {volume} {614}},\ \bibinfo {pages}
  {70} (\bibinfo {year} {2023})}\BibitemShut {NoStop}%
\bibitem [{\citenamefont {Ni}\ \emph {et~al.}(2018)\citenamefont {Ni},
  \citenamefont {Rosenband},\ and\ \citenamefont {Grimes}}]{Ni:2018}%
  \BibitemOpen
  \bibfield  {author} {\bibinfo {author} {\bibfnamefont {K.-K.}\ \bibnamefont
  {Ni}}, \bibinfo {author} {\bibfnamefont {T.}~\bibnamefont {Rosenband}},\ and\
  \bibinfo {author} {\bibfnamefont {D.~D.}\ \bibnamefont {Grimes}},\ }\bibfield
   {title} {\bibinfo {title} {Dipolar exchange quantum logic gate with polar
  molecules},\ }\href {https://doi.org/10.1039/C8SC02355G} {\bibfield
  {journal} {\bibinfo  {journal} {Chem. Sci.}\ }\textbf {\bibinfo {volume}
  {9}},\ \bibinfo {pages} {6830} (\bibinfo {year} {2018})}\BibitemShut
  {NoStop}%
\bibitem [{\citenamefont {Hughes}\ \emph {et~al.}(2020)\citenamefont {Hughes},
  \citenamefont {Frye}, \citenamefont {Sawant}, \citenamefont {Bhole},
  \citenamefont {Jones}, \citenamefont {Cornish}, \citenamefont {Tarbutt},
  \citenamefont {Hutson}, \citenamefont {Jaksch},\ and\ \citenamefont
  {Mur-Petit}}]{Hughes:2020}%
  \BibitemOpen
  \bibfield  {author} {\bibinfo {author} {\bibfnamefont {M.}~\bibnamefont
  {Hughes}}, \bibinfo {author} {\bibfnamefont {M.~D.}\ \bibnamefont {Frye}},
  \bibinfo {author} {\bibfnamefont {R.}~\bibnamefont {Sawant}}, \bibinfo
  {author} {\bibfnamefont {G.}~\bibnamefont {Bhole}}, \bibinfo {author}
  {\bibfnamefont {J.~A.}\ \bibnamefont {Jones}}, \bibinfo {author}
  {\bibfnamefont {S.~L.}\ \bibnamefont {Cornish}}, \bibinfo {author}
  {\bibfnamefont {M.~R.}\ \bibnamefont {Tarbutt}}, \bibinfo {author}
  {\bibfnamefont {J.~M.}\ \bibnamefont {Hutson}}, \bibinfo {author}
  {\bibfnamefont {D.}~\bibnamefont {Jaksch}},\ and\ \bibinfo {author}
  {\bibfnamefont {J.}~\bibnamefont {Mur-Petit}},\ }\bibfield  {title} {\bibinfo
  {title} {Robust entangling gate for polar molecules using magnetic and
  microwave fields},\ }\href {https://doi.org/10.1103/PhysRevA.101.062308}
  {\bibfield  {journal} {\bibinfo  {journal} {Phys. Rev. A}\ }\textbf {\bibinfo
  {volume} {101}},\ \bibinfo {pages} {062308} (\bibinfo {year}
  {2020})}\BibitemShut {NoStop}%
\bibitem [{\citenamefont {Kaufman}\ and\ \citenamefont
  {Ni}(2021)}]{Kaufman:2021}%
  \BibitemOpen
  \bibfield  {author} {\bibinfo {author} {\bibfnamefont {A.~M.}\ \bibnamefont
  {Kaufman}}\ and\ \bibinfo {author} {\bibfnamefont {K.-K.}\ \bibnamefont
  {Ni}},\ }\bibfield  {title} {\bibinfo {title} {Quantum science with optical
  tweezer arrays of ultracold atoms and molecules},\ }\href
  {https://doi.org/10.1038/s41567-021-01357-2} {\bibfield  {journal} {\bibinfo
  {journal} {Nat. Phys.}\ }\textbf {\bibinfo {volume} {17}},\ \bibinfo {pages}
  {1324} (\bibinfo {year} {2021})}\BibitemShut {NoStop}%
\bibitem [{\citenamefont {Altman}\ \emph {et~al.}(2021)\citenamefont {Altman},
  \citenamefont {Brown}, \citenamefont {Carleo}, \citenamefont {Carr},
  \citenamefont {Demler}, \citenamefont {Chin}, \citenamefont {DeMarco},
  \citenamefont {Economou}, \citenamefont {Eriksson}, \citenamefont {Fu},
  \citenamefont {Greiner}, \citenamefont {Hazzard}, \citenamefont {Hulet},
  \citenamefont {Koll\'ar}, \citenamefont {Lev}, \citenamefont {Lukin},
  \citenamefont {Ma}, \citenamefont {Mi}, \citenamefont {Misra}, \citenamefont
  {Monroe}, \citenamefont {Murch}, \citenamefont {Nazario}, \citenamefont {Ni},
  \citenamefont {Potter}, \citenamefont {Roushan}, \citenamefont {Saffman},
  \citenamefont {Schleier-Smith}, \citenamefont {Siddiqi}, \citenamefont
  {Simmonds}, \citenamefont {Singh}, \citenamefont {Spielman}, \citenamefont
  {Temme}, \citenamefont {Weiss}, \citenamefont {Vu\ifmmode \check{c}\else
  \v{c}\fi{}kovi\ifmmode~\acute{c}\else \'{c}\fi{}}, \citenamefont
  {Vuleti\ifmmode~\acute{c}\else \'{c}\fi{}}, \citenamefont {Ye},\ and\
  \citenamefont {Zwierlein}}]{Altman:2021}%
  \BibitemOpen
  \bibfield  {author} {\bibinfo {author} {\bibfnamefont {E.}~\bibnamefont
  {Altman}}, \bibinfo {author} {\bibfnamefont {K.~R.}\ \bibnamefont {Brown}},
  \bibinfo {author} {\bibfnamefont {G.}~\bibnamefont {Carleo}}, \bibinfo
  {author} {\bibfnamefont {L.~D.}\ \bibnamefont {Carr}}, \bibinfo {author}
  {\bibfnamefont {E.}~\bibnamefont {Demler}}, \bibinfo {author} {\bibfnamefont
  {C.}~\bibnamefont {Chin}}, \bibinfo {author} {\bibfnamefont {B.}~\bibnamefont
  {DeMarco}}, \bibinfo {author} {\bibfnamefont {S.~E.}\ \bibnamefont
  {Economou}}, \bibinfo {author} {\bibfnamefont {M.~A.}\ \bibnamefont
  {Eriksson}}, \bibinfo {author} {\bibfnamefont {K.-M.~C.}\ \bibnamefont {Fu}},
  \bibinfo {author} {\bibfnamefont {M.}~\bibnamefont {Greiner}}, \bibinfo
  {author} {\bibfnamefont {K.~R.}\ \bibnamefont {Hazzard}}, \bibinfo {author}
  {\bibfnamefont {R.~G.}\ \bibnamefont {Hulet}}, \bibinfo {author}
  {\bibfnamefont {A.~J.}\ \bibnamefont {Koll\'ar}}, \bibinfo {author}
  {\bibfnamefont {B.~L.}\ \bibnamefont {Lev}}, \bibinfo {author} {\bibfnamefont
  {M.~D.}\ \bibnamefont {Lukin}}, \bibinfo {author} {\bibfnamefont
  {R.}~\bibnamefont {Ma}}, \bibinfo {author} {\bibfnamefont {X.}~\bibnamefont
  {Mi}}, \bibinfo {author} {\bibfnamefont {S.}~\bibnamefont {Misra}}, \bibinfo
  {author} {\bibfnamefont {C.}~\bibnamefont {Monroe}}, \bibinfo {author}
  {\bibfnamefont {K.}~\bibnamefont {Murch}}, \bibinfo {author} {\bibfnamefont
  {Z.}~\bibnamefont {Nazario}}, \bibinfo {author} {\bibfnamefont {K.-K.}\
  \bibnamefont {Ni}}, \bibinfo {author} {\bibfnamefont {A.~C.}\ \bibnamefont
  {Potter}}, \bibinfo {author} {\bibfnamefont {P.}~\bibnamefont {Roushan}},
  \bibinfo {author} {\bibfnamefont {M.}~\bibnamefont {Saffman}}, \bibinfo
  {author} {\bibfnamefont {M.}~\bibnamefont {Schleier-Smith}}, \bibinfo
  {author} {\bibfnamefont {I.}~\bibnamefont {Siddiqi}}, \bibinfo {author}
  {\bibfnamefont {R.}~\bibnamefont {Simmonds}}, \bibinfo {author}
  {\bibfnamefont {M.}~\bibnamefont {Singh}}, \bibinfo {author} {\bibfnamefont
  {I.~B.}\ \bibnamefont {Spielman}}, \bibinfo {author} {\bibfnamefont
  {K.}~\bibnamefont {Temme}}, \bibinfo {author} {\bibfnamefont {D.~S.}\
  \bibnamefont {Weiss}}, \bibinfo {author} {\bibfnamefont {J.}~\bibnamefont
  {Vu\ifmmode \check{c}\else \v{c}\fi{}kovi\ifmmode~\acute{c}\else
  \'{c}\fi{}}}, \bibinfo {author} {\bibfnamefont {V.}~\bibnamefont
  {Vuleti\ifmmode~\acute{c}\else \'{c}\fi{}}}, \bibinfo {author} {\bibfnamefont
  {J.}~\bibnamefont {Ye}},\ and\ \bibinfo {author} {\bibfnamefont
  {M.}~\bibnamefont {Zwierlein}},\ }\bibfield  {title} {\bibinfo {title}
  {Quantum simulators: Architectures and opportunities},\ }\href
  {https://doi.org/10.1103/PRXQuantum.2.017003} {\bibfield  {journal} {\bibinfo
   {journal} {PRX Quantum}\ }\textbf {\bibinfo {volume} {2}},\ \bibinfo {pages}
  {017003} (\bibinfo {year} {2021})}\BibitemShut {NoStop}%
\bibitem [{\citenamefont {Bao}\ \emph {et~al.}(2023)\citenamefont {Bao},
  \citenamefont {Yu}, \citenamefont {Anderegg}, \citenamefont {Chae},
  \citenamefont {Ketterle}, \citenamefont {Ni},\ and\ \citenamefont
  {Doyle}}]{Bao:dipolar:2023}%
  \BibitemOpen
  \bibfield  {author} {\bibinfo {author} {\bibfnamefont {Y.}~\bibnamefont
  {Bao}}, \bibinfo {author} {\bibfnamefont {S.~S.}\ \bibnamefont {Yu}},
  \bibinfo {author} {\bibfnamefont {L.}~\bibnamefont {Anderegg}}, \bibinfo
  {author} {\bibfnamefont {E.}~\bibnamefont {Chae}}, \bibinfo {author}
  {\bibfnamefont {W.}~\bibnamefont {Ketterle}}, \bibinfo {author}
  {\bibfnamefont {K.-K.}\ \bibnamefont {Ni}},\ and\ \bibinfo {author}
  {\bibfnamefont {J.~M.}\ \bibnamefont {Doyle}},\ }\bibfield  {title} {\bibinfo
  {title} {Dipolar spin-exchange and entanglement between molecules in an
  optical tweezer array},\ }\href {https://doi.org/10.1126/science.adf8999}
  {\bibfield  {journal} {\bibinfo  {journal} {Science}\ }\textbf {\bibinfo
  {volume} {382}},\ \bibinfo {pages} {1138} (\bibinfo {year}
  {2023})}\BibitemShut {NoStop}%
\bibitem [{\citenamefont {Chin}\ \emph {et~al.}(2010)\citenamefont {Chin},
  \citenamefont {Grimm}, \citenamefont {Julienne},\ and\ \citenamefont
  {Tiesinga}}]{Chin:RMP:2010}%
  \BibitemOpen
  \bibfield  {author} {\bibinfo {author} {\bibfnamefont {C.}~\bibnamefont
  {Chin}}, \bibinfo {author} {\bibfnamefont {R.}~\bibnamefont {Grimm}},
  \bibinfo {author} {\bibfnamefont {P.~S.}\ \bibnamefont {Julienne}},\ and\
  \bibinfo {author} {\bibfnamefont {E.}~\bibnamefont {Tiesinga}},\ }\bibfield
  {title} {\bibinfo {title} {Feshbach resonances in ultracold gases},\ }\href
  {https://doi.org/10.1103/RevModPhys.82.1225} {\bibfield  {journal} {\bibinfo
  {journal} {Rev. Mod. Phys.}\ }\textbf {\bibinfo {volume} {82}},\ \bibinfo
  {pages} {1225} (\bibinfo {year} {2010})}\BibitemShut {NoStop}%
\bibitem [{\citenamefont {Tanzi}\ \emph {et~al.}(2019)\citenamefont {Tanzi},
  \citenamefont {Lucioni}, \citenamefont {Fam\`a}, \citenamefont {Catani},
  \citenamefont {Fioretti}, \citenamefont {Gabbanini}, \citenamefont {Bisset},
  \citenamefont {Santos},\ and\ \citenamefont {Modugno}}]{Tanzi:2019}%
  \BibitemOpen
  \bibfield  {author} {\bibinfo {author} {\bibfnamefont {L.}~\bibnamefont
  {Tanzi}}, \bibinfo {author} {\bibfnamefont {E.}~\bibnamefont {Lucioni}},
  \bibinfo {author} {\bibfnamefont {F.}~\bibnamefont {Fam\`a}}, \bibinfo
  {author} {\bibfnamefont {J.}~\bibnamefont {Catani}}, \bibinfo {author}
  {\bibfnamefont {A.}~\bibnamefont {Fioretti}}, \bibinfo {author}
  {\bibfnamefont {C.}~\bibnamefont {Gabbanini}}, \bibinfo {author}
  {\bibfnamefont {R.~N.}\ \bibnamefont {Bisset}}, \bibinfo {author}
  {\bibfnamefont {L.}~\bibnamefont {Santos}},\ and\ \bibinfo {author}
  {\bibfnamefont {G.}~\bibnamefont {Modugno}},\ }\bibfield  {title} {\bibinfo
  {title} {Observation of a dipolar quantum gas with metastable supersolid
  properties},\ }\href {https://doi.org/10.1103/PhysRevLett.122.130405}
  {\bibfield  {journal} {\bibinfo  {journal} {Phys. Rev. Lett.}\ }\textbf
  {\bibinfo {volume} {122}},\ \bibinfo {pages} {130405} (\bibinfo {year}
  {2019})}\BibitemShut {NoStop}%
\bibitem [{\citenamefont {Hertkorn}\ \emph {et~al.}(2021)\citenamefont
  {Hertkorn}, \citenamefont {Schmidt}, \citenamefont {Guo}, \citenamefont
  {B\"ottcher}, \citenamefont {Ng}, \citenamefont {Graham}, \citenamefont
  {Uerlings}, \citenamefont {Langen}, \citenamefont {Zwierlein},\ and\
  \citenamefont {Pfau}}]{Hertkorn:2021}%
  \BibitemOpen
  \bibfield  {author} {\bibinfo {author} {\bibfnamefont {J.}~\bibnamefont
  {Hertkorn}}, \bibinfo {author} {\bibfnamefont {J.-N.}\ \bibnamefont
  {Schmidt}}, \bibinfo {author} {\bibfnamefont {M.}~\bibnamefont {Guo}},
  \bibinfo {author} {\bibfnamefont {F.}~\bibnamefont {B\"ottcher}}, \bibinfo
  {author} {\bibfnamefont {K.~S.~H.}\ \bibnamefont {Ng}}, \bibinfo {author}
  {\bibfnamefont {S.~D.}\ \bibnamefont {Graham}}, \bibinfo {author}
  {\bibfnamefont {P.}~\bibnamefont {Uerlings}}, \bibinfo {author}
  {\bibfnamefont {T.}~\bibnamefont {Langen}}, \bibinfo {author} {\bibfnamefont
  {M.}~\bibnamefont {Zwierlein}},\ and\ \bibinfo {author} {\bibfnamefont
  {T.}~\bibnamefont {Pfau}},\ }\bibfield  {title} {\bibinfo {title} {Pattern
  formation in quantum ferrofluids: {F}rom supersolids to superglasses},\
  }\href {https://doi.org/10.1103/PhysRevResearch.3.033125} {\bibfield
  {journal} {\bibinfo  {journal} {Phys. Rev. Res.}\ }\textbf {\bibinfo {volume}
  {3}},\ \bibinfo {pages} {033125} (\bibinfo {year} {2021})}\BibitemShut
  {NoStop}%
\bibitem [{\citenamefont {Patscheider}\ \emph {et~al.}(2022)\citenamefont
  {Patscheider}, \citenamefont {Chomaz}, \citenamefont {Natale}, \citenamefont
  {Petter}, \citenamefont {Mark}, \citenamefont {Baier}, \citenamefont {Yang},
  \citenamefont {Wang}, \citenamefont {Bohn},\ and\ \citenamefont
  {Ferlaino}}]{Patscheider:2022}%
  \BibitemOpen
  \bibfield  {author} {\bibinfo {author} {\bibfnamefont {A.}~\bibnamefont
  {Patscheider}}, \bibinfo {author} {\bibfnamefont {L.}~\bibnamefont {Chomaz}},
  \bibinfo {author} {\bibfnamefont {G.}~\bibnamefont {Natale}}, \bibinfo
  {author} {\bibfnamefont {D.}~\bibnamefont {Petter}}, \bibinfo {author}
  {\bibfnamefont {M.~J.}\ \bibnamefont {Mark}}, \bibinfo {author}
  {\bibfnamefont {S.}~\bibnamefont {Baier}}, \bibinfo {author} {\bibfnamefont
  {B.}~\bibnamefont {Yang}}, \bibinfo {author} {\bibfnamefont {R.~R.~W.}\
  \bibnamefont {Wang}}, \bibinfo {author} {\bibfnamefont {J.~L.}\ \bibnamefont
  {Bohn}},\ and\ \bibinfo {author} {\bibfnamefont {F.}~\bibnamefont
  {Ferlaino}},\ }\bibfield  {title} {\bibinfo {title} {Determination of the
  scattering length of erbium atoms},\ }\href
  {https://doi.org/10.1103/PhysRevA.105.063307} {\bibfield  {journal} {\bibinfo
   {journal} {Phys. Rev. A}\ }\textbf {\bibinfo {volume} {105}},\ \bibinfo
  {pages} {063307} (\bibinfo {year} {2022})}\BibitemShut {NoStop}%
\bibitem [{\citenamefont {K\"{o}hler}\ \emph {et~al.}(2006)\citenamefont
  {K\"{o}hler}, \citenamefont {G\'oral},\ and\ \citenamefont
  {Julienne}}]{Kohler:RMP:2006}%
  \BibitemOpen
  \bibfield  {author} {\bibinfo {author} {\bibfnamefont {T.}~\bibnamefont
  {K\"{o}hler}}, \bibinfo {author} {\bibfnamefont {K.}~\bibnamefont
  {G\'oral}},\ and\ \bibinfo {author} {\bibfnamefont {P.~S.}\ \bibnamefont
  {Julienne}},\ }\bibfield  {title} {\bibinfo {title} {Production of cold
  molecules via magnetically tunable {F}eshbach resonances},\ }\href
  {https://doi.org/10.1103/RevModPhys.78.1311} {\bibfield  {journal} {\bibinfo
  {journal} {Rev. Mod. Phys.}\ }\textbf {\bibinfo {volume} {78}},\ \bibinfo
  {pages} {1311} (\bibinfo {year} {2006})}\BibitemShut {NoStop}%
\bibitem [{\citenamefont {Vitanov}\ \emph {et~al.}(2017)\citenamefont
  {Vitanov}, \citenamefont {Rangelov}, \citenamefont {Shore},\ and\
  \citenamefont {Bergmann}}]{Vitanov:2017}%
  \BibitemOpen
  \bibfield  {author} {\bibinfo {author} {\bibfnamefont {N.~V.}\ \bibnamefont
  {Vitanov}}, \bibinfo {author} {\bibfnamefont {A.~A.}\ \bibnamefont
  {Rangelov}}, \bibinfo {author} {\bibfnamefont {B.~W.}\ \bibnamefont
  {Shore}},\ and\ \bibinfo {author} {\bibfnamefont {K.}~\bibnamefont
  {Bergmann}},\ }\bibfield  {title} {\bibinfo {title} {Stimulated raman
  adiabatic passage in physics, chemistry, and beyond},\ }\href
  {https://doi.org/10.1103/RevModPhys.89.015006} {\bibfield  {journal}
  {\bibinfo  {journal} {Rev. Mod. Phys.}\ }\textbf {\bibinfo {volume} {89}},\
  \bibinfo {pages} {015006} (\bibinfo {year} {2017})}\BibitemShut {NoStop}%
\bibitem [{\citenamefont {Micheli}\ \emph {et~al.}(2006)\citenamefont
  {Micheli}, \citenamefont {Brennen},\ and\ \citenamefont
  {Zoller}}]{Micheli:2006}%
  \BibitemOpen
  \bibfield  {author} {\bibinfo {author} {\bibfnamefont {A.}~\bibnamefont
  {Micheli}}, \bibinfo {author} {\bibfnamefont {G.~K.}\ \bibnamefont
  {Brennen}},\ and\ \bibinfo {author} {\bibfnamefont {P.}~\bibnamefont
  {Zoller}},\ }\bibfield  {title} {\bibinfo {title} {A toolbox for lattice-spin
  models with polar molecules},\ }\href {https://doi.org/10.1038/nphys287}
  {\bibfield  {journal} {\bibinfo  {journal} {Nat. Phys.}\ }\textbf {\bibinfo
  {volume} {2}},\ \bibinfo {pages} {341} (\bibinfo {year} {2006})}\BibitemShut
  {NoStop}%
\bibitem [{\citenamefont {Baranov}\ \emph {et~al.}(2012)\citenamefont
  {Baranov}, \citenamefont {Dalmonte}, \citenamefont {Pupillo},\ and\
  \citenamefont {Zoller}}]{Baranov:2012}%
  \BibitemOpen
  \bibfield  {author} {\bibinfo {author} {\bibfnamefont {M.~A.}\ \bibnamefont
  {Baranov}}, \bibinfo {author} {\bibfnamefont {M.}~\bibnamefont {Dalmonte}},
  \bibinfo {author} {\bibfnamefont {G.}~\bibnamefont {Pupillo}},\ and\ \bibinfo
  {author} {\bibfnamefont {P.}~\bibnamefont {Zoller}},\ }\bibfield  {title}
  {\bibinfo {title} {Condensed matter theory of dipolar quantum gases},\ }\href
  {https://doi.org/10.1021/cr2003568} {\bibfield  {journal} {\bibinfo
  {journal} {Chem. Rev.}\ }\textbf {\bibinfo {volume} {112}},\ \bibinfo {pages}
  {5012} (\bibinfo {year} {2012})}\BibitemShut {NoStop}%
\bibitem [{\citenamefont {Karra}\ \emph {et~al.}(2016)\citenamefont {Karra},
  \citenamefont {Sharma}, \citenamefont {Friedrich}, \citenamefont {Kais},\
  and\ \citenamefont {Herschbach}}]{Karra:2016}%
  \BibitemOpen
  \bibfield  {author} {\bibinfo {author} {\bibfnamefont {M.}~\bibnamefont
  {Karra}}, \bibinfo {author} {\bibfnamefont {K.}~\bibnamefont {Sharma}},
  \bibinfo {author} {\bibfnamefont {B.}~\bibnamefont {Friedrich}}, \bibinfo
  {author} {\bibfnamefont {S.}~\bibnamefont {Kais}},\ and\ \bibinfo {author}
  {\bibfnamefont {D.}~\bibnamefont {Herschbach}},\ }\bibfield  {title}
  {\bibinfo {title} {{Prospects for quantum computing with an array of
  ultracold polar paramagnetic molecules}},\ }\href
  {https://doi.org/10.1063/1.4942928} {\bibfield  {journal} {\bibinfo
  {journal} {J. Chem. Phys.}\ }\textbf {\bibinfo {volume} {144}},\ \bibinfo
  {pages} {094301} (\bibinfo {year} {2016})}\BibitemShut {NoStop}%
\bibitem [{\citenamefont {Safronova}\ \emph {et~al.}(2018)\citenamefont
  {Safronova}, \citenamefont {Budker}, \citenamefont {DeMille}, \citenamefont
  {Kimball}, \citenamefont {Derevianko},\ and\ \citenamefont
  {Clark}}]{Safronova:2018}%
  \BibitemOpen
  \bibfield  {author} {\bibinfo {author} {\bibfnamefont {M.~S.}\ \bibnamefont
  {Safronova}}, \bibinfo {author} {\bibfnamefont {D.}~\bibnamefont {Budker}},
  \bibinfo {author} {\bibfnamefont {D.}~\bibnamefont {DeMille}}, \bibinfo
  {author} {\bibfnamefont {D.~F.~J.}\ \bibnamefont {Kimball}}, \bibinfo
  {author} {\bibfnamefont {A.}~\bibnamefont {Derevianko}},\ and\ \bibinfo
  {author} {\bibfnamefont {C.~W.}\ \bibnamefont {Clark}},\ }\bibfield  {title}
  {\bibinfo {title} {Search for new physics with atoms and molecules},\ }\href
  {https://doi.org/10.1103/RevModPhys.90.025008} {\bibfield  {journal}
  {\bibinfo  {journal} {Rev. Mod. Phys.}\ }\textbf {\bibinfo {volume} {90}},\
  \bibinfo {pages} {025008} (\bibinfo {year} {2018})}\BibitemShut {NoStop}%
\bibitem [{\citenamefont {\.Zuchowski}\ \emph {et~al.}(2010)\citenamefont
  {\.Zuchowski}, \citenamefont {Aldegunde},\ and\ \citenamefont
  {Hutson}}]{Zuchowski:RbSr:2010}%
  \BibitemOpen
  \bibfield  {author} {\bibinfo {author} {\bibfnamefont {P.~S.}\ \bibnamefont
  {\.Zuchowski}}, \bibinfo {author} {\bibfnamefont {J.}~\bibnamefont
  {Aldegunde}},\ and\ \bibinfo {author} {\bibfnamefont {J.~M.}\ \bibnamefont
  {Hutson}},\ }\bibfield  {title} {\bibinfo {title} {Ultracold {RbSr} molecules
  can be formed by magnetoassociation},\ }\href
  {https://doi.org/10.1103/PhysRevLett.105.153201} {\bibfield  {journal}
  {\bibinfo  {journal} {Phys. Rev. Lett.}\ }\textbf {\bibinfo {volume} {105}},\
  \bibinfo {pages} {153201} (\bibinfo {year} {2010})}\BibitemShut {NoStop}%
\bibitem [{\citenamefont {Brue}\ and\ \citenamefont
  {Hutson}(2012)}]{Brue:LiYb:2012}%
  \BibitemOpen
  \bibfield  {author} {\bibinfo {author} {\bibfnamefont {D.~A.}\ \bibnamefont
  {Brue}}\ and\ \bibinfo {author} {\bibfnamefont {J.~M.}\ \bibnamefont
  {Hutson}},\ }\bibfield  {title} {\bibinfo {title} {Magnetically tunable
  {F}eshbach resonances in ultracold {Li-Yb} mixtures},\ }\href
  {https://doi.org/10.1103/PhysRevLett.108.043201} {\bibfield  {journal}
  {\bibinfo  {journal} {Phys. Rev. Lett.}\ }\textbf {\bibinfo {volume} {108}},\
  \bibinfo {pages} {043201} (\bibinfo {year} {2012})}\BibitemShut {NoStop}%
\bibitem [{\citenamefont {Brue}\ and\ \citenamefont
  {Hutson}(2013)}]{Brue:AlkYb:2013}%
  \BibitemOpen
  \bibfield  {author} {\bibinfo {author} {\bibfnamefont {D.~A.}\ \bibnamefont
  {Brue}}\ and\ \bibinfo {author} {\bibfnamefont {J.~M.}\ \bibnamefont
  {Hutson}},\ }\bibfield  {title} {\bibinfo {title} {Prospects of forming
  molecules in {$^2\Sigma$} states by magnetoassociation of alkali-metal atoms
  with {Yb}},\ }\href {https://doi.org/10.1103/PhysRevA.87.052709} {\bibfield
  {journal} {\bibinfo  {journal} {Phys. Rev. A}\ }\textbf {\bibinfo {volume}
  {87}},\ \bibinfo {pages} {052709} (\bibinfo {year} {2013})}\BibitemShut
  {NoStop}%
\bibitem [{\citenamefont {Yang}\ \emph {et~al.}(2019)\citenamefont {Yang},
  \citenamefont {Frye}, \citenamefont {Guttridge}, \citenamefont {Aldegunde},
  \citenamefont {\.Zuchowski}, \citenamefont {Cornish},\ and\ \citenamefont
  {Hutson}}]{Yang:CsYb:2019}%
  \BibitemOpen
  \bibfield  {author} {\bibinfo {author} {\bibfnamefont {B.~C.}\ \bibnamefont
  {Yang}}, \bibinfo {author} {\bibfnamefont {M.~D.}\ \bibnamefont {Frye}},
  \bibinfo {author} {\bibfnamefont {A.}~\bibnamefont {Guttridge}}, \bibinfo
  {author} {\bibfnamefont {J.}~\bibnamefont {Aldegunde}}, \bibinfo {author}
  {\bibfnamefont {P.~S.}\ \bibnamefont {\.Zuchowski}}, \bibinfo {author}
  {\bibfnamefont {S.~L.}\ \bibnamefont {Cornish}},\ and\ \bibinfo {author}
  {\bibfnamefont {J.~M.}\ \bibnamefont {Hutson}},\ }\bibfield  {title}
  {\bibinfo {title} {Magnetic {F}eshbach resonances in ultracold collisions
  between {Cs} and {Yb} atoms},\ }\href
  {https://doi.org/10.1103/PhysRevA.100.022704} {\bibfield  {journal} {\bibinfo
   {journal} {Phys. Rev. A}\ }\textbf {\bibinfo {volume} {100}},\ \bibinfo
  {pages} {022704} (\bibinfo {year} {2019})}\BibitemShut {NoStop}%
\bibitem [{\citenamefont {Barb\'{e}}\ \emph {et~al.}(2018)\citenamefont
  {Barb\'{e}}, \citenamefont {Ciamei}, \citenamefont {Pasquiou}, \citenamefont
  {Reichs\"{o}llner}, \citenamefont {Schreck}, \citenamefont {\.Zuchowski},\
  and\ \citenamefont {Hutson}}]{Barbe:RbSr:2018}%
  \BibitemOpen
  \bibfield  {author} {\bibinfo {author} {\bibfnamefont {V.}~\bibnamefont
  {Barb\'{e}}}, \bibinfo {author} {\bibfnamefont {A.}~\bibnamefont {Ciamei}},
  \bibinfo {author} {\bibfnamefont {B.}~\bibnamefont {Pasquiou}}, \bibinfo
  {author} {\bibfnamefont {L.}~\bibnamefont {Reichs\"{o}llner}}, \bibinfo
  {author} {\bibfnamefont {F.}~\bibnamefont {Schreck}}, \bibinfo {author}
  {\bibfnamefont {P.~S.}\ \bibnamefont {\.Zuchowski}},\ and\ \bibinfo {author}
  {\bibfnamefont {J.~M.}\ \bibnamefont {Hutson}},\ }\bibfield  {title}
  {\bibinfo {title} {Observation of {F}eshbach resonances between alkali and
  closed-shell atoms},\ }\href {https://doi.org/10.1038/s41567-018-0169-x}
  {\bibfield  {journal} {\bibinfo  {journal} {Nat. Phys.}\ }\textbf {\bibinfo
  {volume} {14}},\ \bibinfo {pages} {881} (\bibinfo {year} {2018})}\BibitemShut
  {NoStop}%
\bibitem [{\citenamefont {Green}\ \emph {et~al.}(2020)\citenamefont {Green},
  \citenamefont {Li}, \citenamefont {Toh}, \citenamefont {Tang}, \citenamefont
  {McCormack}, \citenamefont {Li}, \citenamefont {Tiesinga}, \citenamefont
  {Kotochigova},\ and\ \citenamefont {Gupta}}]{Green:LiYb-res:2020}%
  \BibitemOpen
  \bibfield  {author} {\bibinfo {author} {\bibfnamefont {A.}~\bibnamefont
  {Green}}, \bibinfo {author} {\bibfnamefont {H.}~\bibnamefont {Li}}, \bibinfo
  {author} {\bibfnamefont {J.~H.~S.}\ \bibnamefont {Toh}}, \bibinfo {author}
  {\bibfnamefont {X.}~\bibnamefont {Tang}}, \bibinfo {author} {\bibfnamefont
  {K.~C.}\ \bibnamefont {McCormack}}, \bibinfo {author} {\bibfnamefont
  {M.}~\bibnamefont {Li}}, \bibinfo {author} {\bibfnamefont {E.}~\bibnamefont
  {Tiesinga}}, \bibinfo {author} {\bibfnamefont {S.}~\bibnamefont
  {Kotochigova}},\ and\ \bibinfo {author} {\bibfnamefont {S.}~\bibnamefont
  {Gupta}},\ }\bibfield  {title} {\bibinfo {title} {Feshbach resonances in
  p-wave three-body recombination within {F}ermi-{F}ermi mixtures of open-shell
  $^6${Li} and closed-shell $^{173}${Yb} atoms},\ }\href
  {https://doi.org/10.1103/PhysRevX.10.031037} {\bibfield  {journal} {\bibinfo
  {journal} {Phys. Rev. X}\ }\textbf {\bibinfo {volume} {10}},\ \bibinfo
  {pages} {031037} (\bibinfo {year} {2020})}\BibitemShut {NoStop}%
\bibitem [{\citenamefont {Franzen}\ \emph {et~al.}(2022)\citenamefont
  {Franzen}, \citenamefont {Guttridge}, \citenamefont {Wilson}, \citenamefont
  {Segal}, \citenamefont {Frye}, \citenamefont {Hutson},\ and\ \citenamefont
  {Cornish}}]{Franzen:CsYb-res:2022}%
  \BibitemOpen
  \bibfield  {author} {\bibinfo {author} {\bibfnamefont {T.}~\bibnamefont
  {Franzen}}, \bibinfo {author} {\bibfnamefont {A.}~\bibnamefont {Guttridge}},
  \bibinfo {author} {\bibfnamefont {K.~E.}\ \bibnamefont {Wilson}}, \bibinfo
  {author} {\bibfnamefont {J.}~\bibnamefont {Segal}}, \bibinfo {author}
  {\bibfnamefont {M.~D.}\ \bibnamefont {Frye}}, \bibinfo {author}
  {\bibfnamefont {J.~M.}\ \bibnamefont {Hutson}},\ and\ \bibinfo {author}
  {\bibfnamefont {S.~L.}\ \bibnamefont {Cornish}},\ }\bibfield  {title}
  {\bibinfo {title} {Observation of magnetic {F}eshbach resonances between {Cs}
  and $^{173}${Yb}},\ }\href {https://doi.org/10.1103/PhysRevResearch.4.043072}
  {\bibfield  {journal} {\bibinfo  {journal} {Phys. Rev. Res.}\ }\textbf
  {\bibinfo {volume} {4}},\ \bibinfo {pages} {043072} (\bibinfo {year}
  {2022})}\BibitemShut {NoStop}%
\bibitem [{\citenamefont {Borkowski}\ \emph {et~al.}(2023)\citenamefont
  {Borkowski}, \citenamefont {Reichs{\"o}llner}, \citenamefont {Thekkeppatt},
  \citenamefont {Barb{\'e}}, \citenamefont {van Roon}, \citenamefont {van
  Druten},\ and\ \citenamefont {Schreck}}]{Borkowski:2023}%
  \BibitemOpen
  \bibfield  {author} {\bibinfo {author} {\bibfnamefont {M.}~\bibnamefont
  {Borkowski}}, \bibinfo {author} {\bibfnamefont {L.}~\bibnamefont
  {Reichs{\"o}llner}}, \bibinfo {author} {\bibfnamefont {P.}~\bibnamefont
  {Thekkeppatt}}, \bibinfo {author} {\bibfnamefont {V.}~\bibnamefont
  {Barb{\'e}}}, \bibinfo {author} {\bibfnamefont {T.}~\bibnamefont {van Roon}},
  \bibinfo {author} {\bibfnamefont {K.}~\bibnamefont {van Druten}},\ and\
  \bibinfo {author} {\bibfnamefont {F.}~\bibnamefont {Schreck}},\ }\bibfield
  {title} {\bibinfo {title} {Active stabilization of kilogauss magnetic fields
  to the ppm level for magnetoassociation on ultranarrow {F}eshbach
  resonances},\ }\href {https://doi.org/10.1063/5.0143825} {\bibfield
  {journal} {\bibinfo  {journal} {Rev. Sci. Instrum.}\ }\textbf {\bibinfo
  {volume} {94}},\ \bibinfo {pages} {073202} (\bibinfo {year}
  {2023})}\BibitemShut {NoStop}%
\bibitem [{\citenamefont {Truppe}\ \emph {et~al.}(2017)\citenamefont {Truppe},
  \citenamefont {Williams}, \citenamefont {Hambach}, \citenamefont {Caldwell},
  \citenamefont {Fitch}, \citenamefont {Hinds}, \citenamefont {Sauer},\ and\
  \citenamefont {Tarbutt}}]{Truppe:MOT:2017}%
  \BibitemOpen
  \bibfield  {author} {\bibinfo {author} {\bibfnamefont {S.}~\bibnamefont
  {Truppe}}, \bibinfo {author} {\bibfnamefont {H.~J.}\ \bibnamefont
  {Williams}}, \bibinfo {author} {\bibfnamefont {M.}~\bibnamefont {Hambach}},
  \bibinfo {author} {\bibfnamefont {L.}~\bibnamefont {Caldwell}}, \bibinfo
  {author} {\bibfnamefont {N.~J.}\ \bibnamefont {Fitch}}, \bibinfo {author}
  {\bibfnamefont {E.~A.}\ \bibnamefont {Hinds}}, \bibinfo {author}
  {\bibfnamefont {B.~E.}\ \bibnamefont {Sauer}},\ and\ \bibinfo {author}
  {\bibfnamefont {M.~R.}\ \bibnamefont {Tarbutt}},\ }\bibfield  {title}
  {\bibinfo {title} {Molecules cooled below the {D}oppler limit},\ }\href
  {https://doi.org/10.1038/nphys4241} {\bibfield  {journal} {\bibinfo
  {journal} {Nat. Phys.}\ }\textbf {\bibinfo {volume} {13}},\ \bibinfo {pages}
  {1173} (\bibinfo {year} {2017})}\BibitemShut {NoStop}%
\bibitem [{\citenamefont {Anderegg}\ \emph {et~al.}(2018)\citenamefont
  {Anderegg}, \citenamefont {Augenbraun}, \citenamefont {Bao}, \citenamefont
  {Burchesky}, \citenamefont {Cheuk}, \citenamefont {Ketterle},\ and\
  \citenamefont {Doyle}}]{Anderegg:2018}%
  \BibitemOpen
  \bibfield  {author} {\bibinfo {author} {\bibfnamefont {L.}~\bibnamefont
  {Anderegg}}, \bibinfo {author} {\bibfnamefont {B.~L.}\ \bibnamefont
  {Augenbraun}}, \bibinfo {author} {\bibfnamefont {Y.}~\bibnamefont {Bao}},
  \bibinfo {author} {\bibfnamefont {S.}~\bibnamefont {Burchesky}}, \bibinfo
  {author} {\bibfnamefont {L.~W.}\ \bibnamefont {Cheuk}}, \bibinfo {author}
  {\bibfnamefont {W.}~\bibnamefont {Ketterle}},\ and\ \bibinfo {author}
  {\bibfnamefont {J.~M.}\ \bibnamefont {Doyle}},\ }\bibfield  {title} {\bibinfo
  {title} {Laser cooling of optically trapped molecules},\ }\href
  {https://doi.org/10.1038/s41567-018-0191-z} {\bibfield  {journal} {\bibinfo
  {journal} {Nat. Phys.}\ }\textbf {\bibinfo {volume} {14}},\ \bibinfo {pages}
  {890} (\bibinfo {year} {2018})}\BibitemShut {NoStop}%
\bibitem [{\citenamefont {Barry}\ \emph {et~al.}(2014)\citenamefont {Barry},
  \citenamefont {McCarron}, \citenamefont {Norrgard}, \citenamefont
  {Steinecker},\ and\ \citenamefont {DeMille}}]{Barry:2014}%
  \BibitemOpen
  \bibfield  {author} {\bibinfo {author} {\bibfnamefont {J.~F.}\ \bibnamefont
  {Barry}}, \bibinfo {author} {\bibfnamefont {D.~J.}\ \bibnamefont {McCarron}},
  \bibinfo {author} {\bibfnamefont {E.~B.}\ \bibnamefont {Norrgard}}, \bibinfo
  {author} {\bibfnamefont {M.~H.}\ \bibnamefont {Steinecker}},\ and\ \bibinfo
  {author} {\bibfnamefont {D.}~\bibnamefont {DeMille}},\ }\bibfield  {title}
  {\bibinfo {title} {Magneto-optical trapping of a diatomic molecule},\ }\href
  {https://doi.org/{10.1038/nature13634}} {\bibfield  {journal} {\bibinfo
  {journal} {Nature}\ }\textbf {\bibinfo {volume} {512}},\ \bibinfo {pages}
  {286} (\bibinfo {year} {2014})}\BibitemShut {NoStop}%
\bibitem [{\citenamefont {McCarron}\ \emph {et~al.}(2018)\citenamefont
  {McCarron}, \citenamefont {Steinecker}, \citenamefont {Zhu},\ and\
  \citenamefont {DeMille}}]{McCarron:2018}%
  \BibitemOpen
  \bibfield  {author} {\bibinfo {author} {\bibfnamefont {D.~J.}\ \bibnamefont
  {McCarron}}, \bibinfo {author} {\bibfnamefont {M.~H.}\ \bibnamefont
  {Steinecker}}, \bibinfo {author} {\bibfnamefont {Y.}~\bibnamefont {Zhu}},\
  and\ \bibinfo {author} {\bibfnamefont {D.}~\bibnamefont {DeMille}},\
  }\bibfield  {title} {\bibinfo {title} {Magnetic trapping of an ultracold gas
  of polar molecules},\ }\href {https://doi.org/10.1103/PhysRevLett.121.013202}
  {\bibfield  {journal} {\bibinfo  {journal} {Phys. Rev. Lett.}\ }\textbf
  {\bibinfo {volume} {121}},\ \bibinfo {pages} {013202} (\bibinfo {year}
  {2018})}\BibitemShut {NoStop}%
\bibitem [{\citenamefont {Ding}\ \emph {et~al.}(2020)\citenamefont {Ding},
  \citenamefont {Wu}, \citenamefont {Finneran}, \citenamefont {Burau},\ and\
  \citenamefont {Ye}}]{Ding:2020}%
  \BibitemOpen
  \bibfield  {author} {\bibinfo {author} {\bibfnamefont {S.}~\bibnamefont
  {Ding}}, \bibinfo {author} {\bibfnamefont {Y.}~\bibnamefont {Wu}}, \bibinfo
  {author} {\bibfnamefont {I.~A.}\ \bibnamefont {Finneran}}, \bibinfo {author}
  {\bibfnamefont {J.~J.}\ \bibnamefont {Burau}},\ and\ \bibinfo {author}
  {\bibfnamefont {J.}~\bibnamefont {Ye}},\ }\bibfield  {title} {\bibinfo
  {title} {{Sub-Doppler cooling and compressed trapping of YO molecules at
  $\mu$K temperatures}},\ }\href {https://doi.org/10.1103/PhysRevX.10.021049}
  {\bibfield  {journal} {\bibinfo  {journal} {Phys. Rev. X}\ }\textbf {\bibinfo
  {volume} {10}},\ \bibinfo {pages} {021049} (\bibinfo {year}
  {2020})}\BibitemShut {NoStop}%
\bibitem [{\citenamefont {Burau}\ \emph {et~al.}(2023)\citenamefont {Burau},
  \citenamefont {Aggarwal}, \citenamefont {Mehling},\ and\ \citenamefont
  {Ye}}]{Burau:2023}%
  \BibitemOpen
  \bibfield  {author} {\bibinfo {author} {\bibfnamefont {J.~J.}\ \bibnamefont
  {Burau}}, \bibinfo {author} {\bibfnamefont {P.}~\bibnamefont {Aggarwal}},
  \bibinfo {author} {\bibfnamefont {K.}~\bibnamefont {Mehling}},\ and\ \bibinfo
  {author} {\bibfnamefont {J.}~\bibnamefont {Ye}},\ }\bibfield  {title}
  {\bibinfo {title} {Blue-detuned magneto-optical trap of molecules},\ }\href
  {https://doi.org/10.1103/PhysRevLett.130.193401} {\bibfield  {journal}
  {\bibinfo  {journal} {Phys. Rev. Lett.}\ }\textbf {\bibinfo {volume} {130}},\
  \bibinfo {pages} {193401} (\bibinfo {year} {2023})}\BibitemShut {NoStop}%
\bibitem [{\citenamefont {Augenbraun}\ \emph {et~al.}(2023)\citenamefont
  {Augenbraun}, \citenamefont {Anderegg}, \citenamefont {Hallas}, \citenamefont
  {Lasner}, \citenamefont {Vilas},\ and\ \citenamefont
  {Doyle}}]{Augenbraun:2023}%
  \BibitemOpen
  \bibfield  {author} {\bibinfo {author} {\bibfnamefont {B.~L.}\ \bibnamefont
  {Augenbraun}}, \bibinfo {author} {\bibfnamefont {L.}~\bibnamefont
  {Anderegg}}, \bibinfo {author} {\bibfnamefont {C.}~\bibnamefont {Hallas}},
  \bibinfo {author} {\bibfnamefont {Z.~D.}\ \bibnamefont {Lasner}}, \bibinfo
  {author} {\bibfnamefont {N.~B.}\ \bibnamefont {Vilas}},\ and\ \bibinfo
  {author} {\bibfnamefont {J.~M.}\ \bibnamefont {Doyle}},\ }\bibfield  {title}
  {\bibinfo {title} {Direct laser cooling of polyatomic molecules},\ }in\ \href
  {https://doi.org/https://doi.org/10.1016/bs.aamop.2023.04.005} {\emph
  {\bibinfo {booktitle} {Advances in Atomic, Molecular, and Optical
  Physics}}},\ Vol.~\bibinfo {volume} {72},\ \bibinfo {editor} {edited by\
  \bibinfo {editor} {\bibfnamefont {L.~F.}\ \bibnamefont {DiMauro}}, \bibinfo
  {editor} {\bibfnamefont {H.}~\bibnamefont {Perrin}},\ and\ \bibinfo {editor}
  {\bibfnamefont {S.~F.}\ \bibnamefont {Yelin}}}\ (\bibinfo  {publisher}
  {Academic Press},\ \bibinfo {year} {2023})\ pp.\ \bibinfo {pages}
  {89--182}\BibitemShut {NoStop}%
\bibitem [{\citenamefont {Pavlovi\ifmmode~\acute{c}\else \'{c}\fi{}}\ \emph
  {et~al.}(2010)\citenamefont {Pavlovi\ifmmode~\acute{c}\else \'{c}\fi{}},
  \citenamefont {Sadeghpour}, \citenamefont {C\^ot\'e},\ and\ \citenamefont
  {Roos}}]{Pavlovic:2010}%
  \BibitemOpen
  \bibfield  {author} {\bibinfo {author} {\bibfnamefont {Z.}~\bibnamefont
  {Pavlovi\ifmmode~\acute{c}\else \'{c}\fi{}}}, \bibinfo {author}
  {\bibfnamefont {H.~R.}\ \bibnamefont {Sadeghpour}}, \bibinfo {author}
  {\bibfnamefont {R.}~\bibnamefont {C\^ot\'e}},\ and\ \bibinfo {author}
  {\bibfnamefont {B.~O.}\ \bibnamefont {Roos}},\ }\bibfield  {title} {\bibinfo
  {title} {{CrRb}: A molecule with large magnetic and electric dipole
  moments},\ }\href {https://doi.org/10.1103/PhysRevA.81.052706} {\bibfield
  {journal} {\bibinfo  {journal} {Phys. Rev. A}\ }\textbf {\bibinfo {volume}
  {81}},\ \bibinfo {pages} {052706} (\bibinfo {year} {2010})}\BibitemShut
  {NoStop}%
\bibitem [{\citenamefont {Tomza}(2013)}]{Tomza:2013}%
  \BibitemOpen
  \bibfield  {author} {\bibinfo {author} {\bibfnamefont {M.}~\bibnamefont
  {Tomza}},\ }\bibfield  {title} {\bibinfo {title} {Prospects for ultracold
  polar and magnetic chromium--closed-shell-atom molecules},\ }\href
  {https://doi.org/10.1103/PhysRevA.88.012519} {\bibfield  {journal} {\bibinfo
  {journal} {Phys. Rev. A}\ }\textbf {\bibinfo {volume} {88}},\ \bibinfo
  {pages} {012519} (\bibinfo {year} {2013})}\BibitemShut {NoStop}%
\bibitem [{\citenamefont {Tomza}(2014)}]{Tomza:2014}%
  \BibitemOpen
  \bibfield  {author} {\bibinfo {author} {\bibfnamefont {M.}~\bibnamefont
  {Tomza}},\ }\bibfield  {title} {\bibinfo {title} {Ab initio properties of the
  ground-state polar and paramagnetic europium--alkali-metal-atom and
  europium--alkaline-earth-metal-atom molecules},\ }\href
  {https://doi.org/10.1103/PhysRevA.90.022514} {\bibfield  {journal} {\bibinfo
  {journal} {Phys. Rev. A}\ }\textbf {\bibinfo {volume} {90}},\ \bibinfo
  {pages} {022514} (\bibinfo {year} {2014})}\BibitemShut {NoStop}%
\bibitem [{\citenamefont {Zaremba-Kopczyk}\ \emph {et~al.}(2018)\citenamefont
  {Zaremba-Kopczyk}, \citenamefont {\ifmmode~\dot{Z}\else \.{Z}\fi{}uchowski},\
  and\ \citenamefont {Tomza}}]{Zaremba:2018}%
  \BibitemOpen
  \bibfield  {author} {\bibinfo {author} {\bibfnamefont {K.}~\bibnamefont
  {Zaremba-Kopczyk}}, \bibinfo {author} {\bibfnamefont {P.~S.}\ \bibnamefont
  {\ifmmode~\dot{Z}\else \.{Z}\fi{}uchowski}},\ and\ \bibinfo {author}
  {\bibfnamefont {M.}~\bibnamefont {Tomza}},\ }\bibfield  {title} {\bibinfo
  {title} {Magnetically tunable {F}eshbach resonances in ultracold gases of
  europium atoms and mixtures of europium and alkali-metal atoms},\ }\href
  {https://doi.org/10.1103/PhysRevA.98.032704} {\bibfield  {journal} {\bibinfo
  {journal} {Phys. Rev. A}\ }\textbf {\bibinfo {volume} {98}},\ \bibinfo
  {pages} {032704} (\bibinfo {year} {2018})}\BibitemShut {NoStop}%
\bibitem [{\citenamefont {Kosicki}\ \emph {et~al.}(2020)\citenamefont
  {Kosicki}, \citenamefont {Borkowski},\ and\ \citenamefont
  {{\.Z}uchowski}}]{Kosicki:ErYb:2020}%
  \BibitemOpen
  \bibfield  {author} {\bibinfo {author} {\bibfnamefont {M.~B.}\ \bibnamefont
  {Kosicki}}, \bibinfo {author} {\bibfnamefont {M.}~\bibnamefont {Borkowski}},\
  and\ \bibinfo {author} {\bibfnamefont {P.~S.}\ \bibnamefont
  {{\.Z}uchowski}},\ }\bibfield  {title} {\bibinfo {title} {Quantum chaos in
  {F}eshbach resonances of the {ErYb} system},\ }\href
  {https://doi.org/10.1088/1367-2630/ab6c36} {\bibfield  {journal} {\bibinfo
  {journal} {New J. Phys.}\ }\textbf {\bibinfo {volume} {22}},\ \bibinfo
  {pages} {023024} (\bibinfo {year} {2020})}\BibitemShut {NoStop}%
\bibitem [{\citenamefont {Frye}\ \emph {et~al.}(2020)\citenamefont {Frye},
  \citenamefont {Cornish},\ and\ \citenamefont {Hutson}}]{Frye:DyYb:2020}%
  \BibitemOpen
  \bibfield  {author} {\bibinfo {author} {\bibfnamefont {M.~D.}\ \bibnamefont
  {Frye}}, \bibinfo {author} {\bibfnamefont {S.~L.}\ \bibnamefont {Cornish}},\
  and\ \bibinfo {author} {\bibfnamefont {J.~M.}\ \bibnamefont {Hutson}},\
  }\bibfield  {title} {\bibinfo {title} {Prospects of forming high-spin polar
  molecules from ultracold atoms},\ }\href
  {https://doi.org/10.1103/PhysRevX.10.041005} {\bibfield  {journal} {\bibinfo
  {journal} {Phys. Rev. X}\ }\textbf {\bibinfo {volume} {10}},\ \bibinfo
  {pages} {041005} (\bibinfo {year} {2020})}\BibitemShut {NoStop}%
\bibitem [{\citenamefont {Ciamei}\ \emph {et~al.}(2022)\citenamefont {Ciamei},
  \citenamefont {Finelli}, \citenamefont {Trenkwalder}, \citenamefont
  {Inguscio}, \citenamefont {Simoni},\ and\ \citenamefont
  {Zaccanti}}]{Ciamei:2022}%
  \BibitemOpen
  \bibfield  {author} {\bibinfo {author} {\bibfnamefont {A.}~\bibnamefont
  {Ciamei}}, \bibinfo {author} {\bibfnamefont {S.}~\bibnamefont {Finelli}},
  \bibinfo {author} {\bibfnamefont {A.}~\bibnamefont {Trenkwalder}}, \bibinfo
  {author} {\bibfnamefont {M.}~\bibnamefont {Inguscio}}, \bibinfo {author}
  {\bibfnamefont {A.}~\bibnamefont {Simoni}},\ and\ \bibinfo {author}
  {\bibfnamefont {M.}~\bibnamefont {Zaccanti}},\ }\bibfield  {title} {\bibinfo
  {title} {Exploring ultracold collisions in
  $^{6}\mathrm{Li}\text{\ensuremath{-}}^{53}\mathrm{Cr}$ fermi mixtures:
  Feshbach resonances and scattering properties of a novel alkali-transition
  metal system},\ }\href {https://doi.org/10.1103/PhysRevLett.129.093402}
  {\bibfield  {journal} {\bibinfo  {journal} {Phys. Rev. Lett.}\ }\textbf
  {\bibinfo {volume} {129}},\ \bibinfo {pages} {093402} (\bibinfo {year}
  {2022})}\BibitemShut {NoStop}%
\bibitem [{\citenamefont {Soave}\ \emph {et~al.}(2023)\citenamefont {Soave},
  \citenamefont {Canali}, \citenamefont {Ye}, \citenamefont {Kreyer},
  \citenamefont {Kirilov},\ and\ \citenamefont {Grimm}}]{Soave:DyK:2023}%
  \BibitemOpen
  \bibfield  {author} {\bibinfo {author} {\bibfnamefont {E.}~\bibnamefont
  {Soave}}, \bibinfo {author} {\bibfnamefont {A.}~\bibnamefont {Canali}},
  \bibinfo {author} {\bibfnamefont {Z.-X.}\ \bibnamefont {Ye}}, \bibinfo
  {author} {\bibfnamefont {M.}~\bibnamefont {Kreyer}}, \bibinfo {author}
  {\bibfnamefont {E.}~\bibnamefont {Kirilov}},\ and\ \bibinfo {author}
  {\bibfnamefont {R.}~\bibnamefont {Grimm}},\ }\bibfield  {title} {\bibinfo
  {title} {Optically trapped {F}eshbach molecules of fermionic
  $^{161}\mathrm{Dy}$ and $^{40}\mathrm{K}$},\ }\href
  {https://doi.org/10.1103/PhysRevResearch.5.033117} {\bibfield  {journal}
  {\bibinfo  {journal} {Phys. Rev. Res.}\ }\textbf {\bibinfo {volume} {5}},\
  \bibinfo {pages} {033117} (\bibinfo {year} {2023})}\BibitemShut {NoStop}%
\bibitem [{\citenamefont {Finelli}\ \emph {et~al.}(2024)\citenamefont
  {Finelli}, \citenamefont {Ciamei}, \citenamefont {Restivo}, \citenamefont
  {Schemmer}, \citenamefont {Cosco}, \citenamefont {Inguscio}, \citenamefont
  {Trenkwalder}, \citenamefont {Zaremba-Kopczyk}, \citenamefont {Gronowski},
  \citenamefont {Tomza},\ and\ \citenamefont {Zaccanti}}]{Finelli:LiCr:2024}%
  \BibitemOpen
  \bibfield  {author} {\bibinfo {author} {\bibfnamefont {S.}~\bibnamefont
  {Finelli}}, \bibinfo {author} {\bibfnamefont {A.}~\bibnamefont {Ciamei}},
  \bibinfo {author} {\bibfnamefont {B.}~\bibnamefont {Restivo}}, \bibinfo
  {author} {\bibfnamefont {M.}~\bibnamefont {Schemmer}}, \bibinfo {author}
  {\bibfnamefont {A.}~\bibnamefont {Cosco}}, \bibinfo {author} {\bibfnamefont
  {M.}~\bibnamefont {Inguscio}}, \bibinfo {author} {\bibfnamefont
  {A.}~\bibnamefont {Trenkwalder}}, \bibinfo {author} {\bibfnamefont
  {K.}~\bibnamefont {Zaremba-Kopczyk}}, \bibinfo {author} {\bibfnamefont
  {M.}~\bibnamefont {Gronowski}}, \bibinfo {author} {\bibfnamefont
  {M.}~\bibnamefont {Tomza}},\ and\ \bibinfo {author} {\bibfnamefont
  {M.}~\bibnamefont {Zaccanti}},\ }\bibfield  {title} {\bibinfo {title}
  {Ultracold {LiCr}: a new pathway to quantum gases of paramagnetic polar
  molecules},\ }\href@noop {} {\bibfield  {journal} {\bibinfo  {journal} {arXiv
  preprint arXiv:2402.08337}\ } (\bibinfo {year} {2024})}\BibitemShut {NoStop}%
\bibitem [{\citenamefont {Brown}\ and\ \citenamefont
  {Carrington}(2003)}]{Brown:2003}%
  \BibitemOpen
  \bibfield  {author} {\bibinfo {author} {\bibfnamefont {J.~M.}\ \bibnamefont
  {Brown}}\ and\ \bibinfo {author} {\bibfnamefont {A.}~\bibnamefont
  {Carrington}},\ }\href@noop {} {\emph {\bibinfo {title} {Rotational
  Spectroscopy of Diatomic Molecules}}}\ (\bibinfo  {publisher} {Cambridge
  University Press},\ \bibinfo {address} {Cambridge},\ \bibinfo {year}
  {2003})\BibitemShut {NoStop}%
\bibitem [{\citenamefont {Stoof}\ \emph {et~al.}(1988)\citenamefont {Stoof},
  \citenamefont {Koelman},\ and\ \citenamefont {Verhaar}}]{Stoof:1988}%
  \BibitemOpen
  \bibfield  {author} {\bibinfo {author} {\bibfnamefont {H.~T.~C.}\
  \bibnamefont {Stoof}}, \bibinfo {author} {\bibfnamefont {J.~M. V.~A.}\
  \bibnamefont {Koelman}},\ and\ \bibinfo {author} {\bibfnamefont {B.~J.}\
  \bibnamefont {Verhaar}},\ }\bibfield  {title} {\bibinfo {title}
  {Spin-exchange and dipole relaxation rates in atomic hydrogen: rigorous and
  simplified calculations},\ }\href {https://doi.org/10.1103/PhysRevB.38.4688}
  {\bibfield  {journal} {\bibinfo  {journal} {Phys. Rev. B}\ }\textbf {\bibinfo
  {volume} {38}},\ \bibinfo {pages} {4688} (\bibinfo {year}
  {1988})}\BibitemShut {NoStop}%
\bibitem [{\citenamefont {Mies}\ \emph {et~al.}(1996)\citenamefont {Mies},
  \citenamefont {Williams}, \citenamefont {Julienne},\ and\ \citenamefont
  {Krauss}}]{Mies:1996}%
  \BibitemOpen
  \bibfield  {author} {\bibinfo {author} {\bibfnamefont {F.~H.}\ \bibnamefont
  {Mies}}, \bibinfo {author} {\bibfnamefont {C.~J.}\ \bibnamefont {Williams}},
  \bibinfo {author} {\bibfnamefont {P.~S.}\ \bibnamefont {Julienne}},\ and\
  \bibinfo {author} {\bibfnamefont {M.}~\bibnamefont {Krauss}},\ }\bibfield
  {title} {\bibinfo {title} {Estimating bounds on collisional relaxation rates
  of spin-polarized $^{87}${Rb} atoms at ultracold temperatures},\ }\href
  {https://doi.org/10.6028/jres.101.052} {\bibfield  {journal} {\bibinfo
  {journal} {J. Res. Natl. Inst. Stand. Technol.}\ }\textbf {\bibinfo {volume}
  {101}},\ \bibinfo {pages} {521} (\bibinfo {year} {1996})}\BibitemShut
  {NoStop}%
\bibitem [{\citenamefont {Hutson}\ \emph {et~al.}(2008)\citenamefont {Hutson},
  \citenamefont {Tiesinga},\ and\ \citenamefont {Julienne}}]{Hutson:Cs2:2008}%
  \BibitemOpen
  \bibfield  {author} {\bibinfo {author} {\bibfnamefont {J.~M.}\ \bibnamefont
  {Hutson}}, \bibinfo {author} {\bibfnamefont {E.}~\bibnamefont {Tiesinga}},\
  and\ \bibinfo {author} {\bibfnamefont {P.~S.}\ \bibnamefont {Julienne}},\
  }\bibfield  {title} {\bibinfo {title} {Avoided crossings between bound states
  of ultracold cesium dimers},\ }\href
  {https://doi.org/10.1103/PhysRevA.78.052703} {\bibfield  {journal} {\bibinfo
  {journal} {Phys. Rev. A}\ }\textbf {\bibinfo {volume} {78}},\ \bibinfo
  {pages} {052703} (\bibinfo {year} {2008})}\BibitemShut {NoStop}%
\bibitem [{Note1()}]{Note1}%
  \BibitemOpen
  \bibinfo {note} {The strength of this coupling is sometimes described by the
  so-called zero-field splitting parameter $D$, particularly in the electron
  paramagnetic resonance community; this is related by $\lambda
  =D/2$.}\BibitemShut {Stop}%
\bibitem [{\citenamefont {Vahtras}\ \emph {et~al.}(2002)\citenamefont
  {Vahtras}, \citenamefont {Loboda}, \citenamefont {Minaev}, \citenamefont
  {{\AA}gren},\ and\ \citenamefont {Ruud}}]{Vahtras:2002}%
  \BibitemOpen
  \bibfield  {author} {\bibinfo {author} {\bibfnamefont {O.}~\bibnamefont
  {Vahtras}}, \bibinfo {author} {\bibfnamefont {O.}~\bibnamefont {Loboda}},
  \bibinfo {author} {\bibfnamefont {B.}~\bibnamefont {Minaev}}, \bibinfo
  {author} {\bibfnamefont {H.}~\bibnamefont {{\AA}gren}},\ and\ \bibinfo
  {author} {\bibfnamefont {K.}~\bibnamefont {Ruud}},\ }\bibfield  {title}
  {\bibinfo {title} {Ab initio calculations of zero-field splitting
  parameters},\ }\href
  {https://doi.org/https://doi.org/10.1016/S0301-0104(02)00451-2} {\bibfield
  {journal} {\bibinfo  {journal} {Chem. Phys.}\ }\textbf {\bibinfo {volume}
  {279}},\ \bibinfo {pages} {133} (\bibinfo {year} {2002})}\BibitemShut
  {NoStop}%
\bibitem [{\citenamefont {Neese}(2007)}]{Neese:ZFS:2007}%
  \BibitemOpen
  \bibfield  {author} {\bibinfo {author} {\bibfnamefont {F.}~\bibnamefont
  {Neese}},\ }\bibfield  {title} {\bibinfo {title} {{Calculation of the
  zero-field splitting tensor on the basis of hybrid density functional and
  Hartree-Fock theory}},\ }\href {https://doi.org/10.1063/1.2772857} {\bibfield
   {journal} {\bibinfo  {journal} {J. Chem. Phys.}\ }\textbf {\bibinfo {volume}
  {127}},\ \bibinfo {pages} {164112} (\bibinfo {year} {2007})}\BibitemShut
  {NoStop}%
\bibitem [{\citenamefont {Te~Velde}\ \emph {et~al.}(2001)\citenamefont
  {Te~Velde}, \citenamefont {Bickelhaupt}, \citenamefont {Baerends},
  \citenamefont {Fonseca~Guerra}, \citenamefont {van Gisbergen}, \citenamefont
  {Snijders},\ and\ \citenamefont {Ziegler}}]{TeVelde:ADF:2001}%
  \BibitemOpen
  \bibfield  {author} {\bibinfo {author} {\bibfnamefont {G.~t.}\ \bibnamefont
  {Te~Velde}}, \bibinfo {author} {\bibfnamefont {F.~M.}\ \bibnamefont
  {Bickelhaupt}}, \bibinfo {author} {\bibfnamefont {E.~J.}\ \bibnamefont
  {Baerends}}, \bibinfo {author} {\bibfnamefont {C.}~\bibnamefont
  {Fonseca~Guerra}}, \bibinfo {author} {\bibfnamefont {S.~J.}\ \bibnamefont
  {van Gisbergen}}, \bibinfo {author} {\bibfnamefont {J.~G.}\ \bibnamefont
  {Snijders}},\ and\ \bibinfo {author} {\bibfnamefont {T.}~\bibnamefont
  {Ziegler}},\ }\bibfield  {title} {\bibinfo {title} {Chemistry with {ADF}},\
  }\href {https://doi.org/10.1002/jcc.1056} {\bibfield  {journal} {\bibinfo
  {journal} {J. Comp. Chem.}\ }\textbf {\bibinfo {volume} {22}},\ \bibinfo
  {pages} {931} (\bibinfo {year} {2001})}\BibitemShut {NoStop}%
\bibitem [{ADF(2023)}]{ADF}%
  \BibitemOpen
  \href@noop {} {\bibinfo {title} {Adf 2023.1}},\ \bibinfo {howpublished} {SCM,
  Theoretical Chemistry, Vrije Universiteit, Amsterdam, The Netherlands,
  http://www.scm.com} (\bibinfo {year} {2023})\BibitemShut {NoStop}%
\bibitem [{\citenamefont {van Lenthe}\ \emph {et~al.}(1994)\citenamefont {van
  Lenthe}, \citenamefont {Baerends},\ and\ \citenamefont
  {Snijders}}]{vanLenthe:1994}%
  \BibitemOpen
  \bibfield  {author} {\bibinfo {author} {\bibfnamefont {E.}~\bibnamefont {van
  Lenthe}}, \bibinfo {author} {\bibfnamefont {E.-J.}\ \bibnamefont
  {Baerends}},\ and\ \bibinfo {author} {\bibfnamefont {J.~G.}\ \bibnamefont
  {Snijders}},\ }\bibfield  {title} {\bibinfo {title} {Relativistic total
  energy using regular approximations},\ }\href
  {https://doi.org/10.1063/1.467943} {\bibfield  {journal} {\bibinfo  {journal}
  {J. Chem. Phys.}\ }\textbf {\bibinfo {volume} {101}},\ \bibinfo {pages}
  {9783} (\bibinfo {year} {1994})}\BibitemShut {NoStop}%
\bibitem [{\citenamefont {Van~Lenthe}\ and\ \citenamefont
  {Baerends}(2003)}]{LentheJCC03}%
  \BibitemOpen
  \bibfield  {author} {\bibinfo {author} {\bibfnamefont {E.}~\bibnamefont
  {Van~Lenthe}}\ and\ \bibinfo {author} {\bibfnamefont {E.~J.}\ \bibnamefont
  {Baerends}},\ }\bibfield  {title} {\bibinfo {title} {Optimized {S}later-type
  basis sets for the elements 1--118},\ }\href
  {https://doi.org/10.1002/jcc.10255} {\bibfield  {journal} {\bibinfo
  {journal} {J. Comp. Chem.}\ }\textbf {\bibinfo {volume} {24}},\ \bibinfo
  {pages} {1142} (\bibinfo {year} {2003})}\BibitemShut {NoStop}%
\bibitem [{\citenamefont {Perdew}\ \emph {et~al.}(1996)\citenamefont {Perdew},
  \citenamefont {Burke},\ and\ \citenamefont {Ernzerhof}}]{Perdew:1996}%
  \BibitemOpen
  \bibfield  {author} {\bibinfo {author} {\bibfnamefont {J.~P.}\ \bibnamefont
  {Perdew}}, \bibinfo {author} {\bibfnamefont {K.}~\bibnamefont {Burke}},\ and\
  \bibinfo {author} {\bibfnamefont {M.}~\bibnamefont {Ernzerhof}},\ }\bibfield
  {title} {\bibinfo {title} {Generalized gradient approximation made simple},\
  }\href {https://doi.org/10.1103/PhysRevLett.77.3865} {\bibfield  {journal}
  {\bibinfo  {journal} {Phys. Rev. Lett.}\ }\textbf {\bibinfo {volume} {77}},\
  \bibinfo {pages} {3865} (\bibinfo {year} {1996})}\BibitemShut {NoStop}%
\bibitem [{\citenamefont {{Le Roy}}\ and\ \citenamefont
  {Henderson}(2007)}]{LeRoy:2007}%
  \BibitemOpen
  \bibfield  {author} {\bibinfo {author} {\bibfnamefont {R.~J.}\ \bibnamefont
  {{Le Roy}}}\ and\ \bibinfo {author} {\bibfnamefont {R.~D.~E.}\ \bibnamefont
  {Henderson}},\ }\bibfield  {title} {\bibinfo {title} {A new potential
  function form incorporating extended long-range behaviour: application to
  ground-state {Ca}$_2$},\ }\href {https://doi.org/10.1080/00268970701241656}
  {\bibfield  {journal} {\bibinfo  {journal} {Mol. Phys.}\ }\textbf {\bibinfo
  {volume} {105}},\ \bibinfo {pages} {663} (\bibinfo {year}
  {2007})}\BibitemShut {NoStop}%
\bibitem [{sup()}]{supp_mat}%
  \BibitemOpen
  \href@noop {} {\bibinfo {title} {See supplemental material at [url will be
  inserted by publisher] for further details of the interaction potential,
  coupling, coupled-channels calculations, and full tabulations of all
  resonance parameters.}}\BibitemShut {Stop}%
\bibitem [{\citenamefont {Alexander}\ and\ \citenamefont
  {Manolopoulos}(1987)}]{Alexander:1987}%
  \BibitemOpen
  \bibfield  {author} {\bibinfo {author} {\bibfnamefont {M.~H.}\ \bibnamefont
  {Alexander}}\ and\ \bibinfo {author} {\bibfnamefont {D.~E.}\ \bibnamefont
  {Manolopoulos}},\ }\bibfield  {title} {\bibinfo {title} {A stable linear
  reference potential algorithm for solution of the quantum close-coupled
  equations in molecular scattering theory},\ }\href
  {https://doi.org/10.1063/1.452154} {\bibfield  {journal} {\bibinfo  {journal}
  {J. Chem. Phys.}\ }\textbf {\bibinfo {volume} {86}},\ \bibinfo {pages} {2044}
  (\bibinfo {year} {1987})}\BibitemShut {NoStop}%
\bibitem [{\citenamefont {Manolopoulos}\ and\ \citenamefont
  {Gray}(1995)}]{MG:symplectic:1995}%
  \BibitemOpen
  \bibfield  {author} {\bibinfo {author} {\bibfnamefont {D.~E.}\ \bibnamefont
  {Manolopoulos}}\ and\ \bibinfo {author} {\bibfnamefont {S.~K.}\ \bibnamefont
  {Gray}},\ }\bibfield  {title} {\bibinfo {title} {Symplectic integrators for
  the multichannel {S}chr{\"o}dinger equation},\ }\href
  {https://doi.org/10.1063/1.468871} {\bibfield  {journal} {\bibinfo  {journal}
  {J. Chem. Phys.}\ }\textbf {\bibinfo {volume} {102}},\ \bibinfo {pages}
  {9214} (\bibinfo {year} {1995})}\BibitemShut {NoStop}%
\bibitem [{\citenamefont {Calvo}\ and\ \citenamefont {Sanz-Serna}(1993)}]{CS4}%
  \BibitemOpen
  \bibfield  {author} {\bibinfo {author} {\bibfnamefont {M.~P.}\ \bibnamefont
  {Calvo}}\ and\ \bibinfo {author} {\bibfnamefont {J.~M.}\ \bibnamefont
  {Sanz-Serna}},\ }\bibfield  {title} {\bibinfo {title} {The development of
  variable-step symplectic integrators, with application to the two-body
  problem},\ }\href {https://doi.org/10.1137/0914057} {\bibfield  {journal}
  {\bibinfo  {journal} {{SIAM} J. Sci. Comput.}\ }\textbf {\bibinfo {volume}
  {14}},\ \bibinfo {pages} {936} (\bibinfo {year} {1993})}\BibitemShut
  {NoStop}%
\bibitem [{\citenamefont {Hutson}\ and\ \citenamefont
  {Le~Sueur}(2019{\natexlab{a}})}]{molscat:2019}%
  \BibitemOpen
  \bibfield  {author} {\bibinfo {author} {\bibfnamefont {J.~M.}\ \bibnamefont
  {Hutson}}\ and\ \bibinfo {author} {\bibfnamefont {C.~R.}\ \bibnamefont
  {Le~Sueur}},\ }\bibfield  {title} {\bibinfo {title} {{\sc molscat}: a program
  for non-reactive quantum scattering calculations on atomic and molecular
  collisions},\ }\href {https://doi.org/doi:10.1016/j.cpc.2019.02.014}
  {\bibfield  {journal} {\bibinfo  {journal} {Comp. Phys. Comm.}\ }\textbf
  {\bibinfo {volume} {241}},\ \bibinfo {pages} {9} (\bibinfo {year}
  {2019}{\natexlab{a}})}\BibitemShut {NoStop}%
\bibitem [{\citenamefont {Hutson}\ and\ \citenamefont
  {Le~Sueur}(2022)}]{mbf-github:2022}%
  \BibitemOpen
  \bibfield  {author} {\bibinfo {author} {\bibfnamefont {J.~M.}\ \bibnamefont
  {Hutson}}\ and\ \bibinfo {author} {\bibfnamefont {C.~R.}\ \bibnamefont
  {Le~Sueur}},\ }\href@noop {} {\bibinfo {title} {{\sc molscat}, {\sc bound}
  and {\sc field}, version 2022.0}},\ \bibinfo {howpublished}
  {\url{https://github.com/molscat/molscat}} (\bibinfo {year}
  {2022})\BibitemShut {NoStop}%
\bibitem [{\citenamefont {Hutson}\ and\ \citenamefont
  {Le~Sueur}(2019{\natexlab{b}})}]{bound+field:2019}%
  \BibitemOpen
  \bibfield  {author} {\bibinfo {author} {\bibfnamefont {J.~M.}\ \bibnamefont
  {Hutson}}\ and\ \bibinfo {author} {\bibfnamefont {C.~R.}\ \bibnamefont
  {Le~Sueur}},\ }\bibfield  {title} {\bibinfo {title} {{\sc bound} and {\sc
  field}: programs for calculating bound states of interacting pairs of atoms
  and molecules},\ }\href {https://doi.org/doi:10.1016/j.cpc.2019.02.017}
  {\bibfield  {journal} {\bibinfo  {journal} {Comp. Phys. Comm.}\ }\textbf
  {\bibinfo {volume} {241}},\ \bibinfo {pages} {1} (\bibinfo {year}
  {2019}{\natexlab{b}})}\BibitemShut {NoStop}%
\bibitem [{\citenamefont {Moerdijk}\ \emph {et~al.}(1995)\citenamefont
  {Moerdijk}, \citenamefont {Verhaar},\ and\ \citenamefont
  {Axelsson}}]{Moerdijk:1995}%
  \BibitemOpen
  \bibfield  {author} {\bibinfo {author} {\bibfnamefont {A.~J.}\ \bibnamefont
  {Moerdijk}}, \bibinfo {author} {\bibfnamefont {B.~J.}\ \bibnamefont
  {Verhaar}},\ and\ \bibinfo {author} {\bibfnamefont {A.}~\bibnamefont
  {Axelsson}},\ }\bibfield  {title} {\bibinfo {title} {Resonances in ultracold
  collisions of $^6${Li}, $^7${Li}, and $^{23}${Na}},\ }\href
  {https://doi.org/10.1103/PhysRevA.51.4852} {\bibfield  {journal} {\bibinfo
  {journal} {Phys. Rev. A}\ }\textbf {\bibinfo {volume} {51}},\ \bibinfo
  {pages} {4852} (\bibinfo {year} {1995})}\BibitemShut {NoStop}%
\bibitem [{\citenamefont {Gribakin}\ and\ \citenamefont
  {Flambaum}(1993)}]{Gribakin:1993}%
  \BibitemOpen
  \bibfield  {author} {\bibinfo {author} {\bibfnamefont {G.~F.}\ \bibnamefont
  {Gribakin}}\ and\ \bibinfo {author} {\bibfnamefont {V.~V.}\ \bibnamefont
  {Flambaum}},\ }\bibfield  {title} {\bibinfo {title} {Calculation of the
  scattering length in atomic collisions using the semiclassical
  approximation},\ }\href {https://doi.org/10.1103/PhysRevA.48.546} {\bibfield
  {journal} {\bibinfo  {journal} {Phys. Rev. A}\ }\textbf {\bibinfo {volume}
  {48}},\ \bibinfo {pages} {546} (\bibinfo {year} {1993})}\BibitemShut
  {NoStop}%
\bibitem [{\citenamefont {Frye}\ and\ \citenamefont
  {Hutson}(2017)}]{Frye:resonance:2017}%
  \BibitemOpen
  \bibfield  {author} {\bibinfo {author} {\bibfnamefont {M.~D.}\ \bibnamefont
  {Frye}}\ and\ \bibinfo {author} {\bibfnamefont {J.~M.}\ \bibnamefont
  {Hutson}},\ }\bibfield  {title} {\bibinfo {title} {Characterizing {F}eshbach
  resonances in ultracold scattering calculations},\ }\href
  {https://doi.org/10.1103/PhysRevA.96.042705} {\bibfield  {journal} {\bibinfo
  {journal} {Phys. Rev. A}\ }\textbf {\bibinfo {volume} {96}},\ \bibinfo
  {pages} {042705} (\bibinfo {year} {2017})}\BibitemShut {NoStop}%
\bibitem [{\citenamefont {Frye}\ and\ \citenamefont
  {Hutson}(2020)}]{Frye:quasibound:2020}%
  \BibitemOpen
  \bibfield  {author} {\bibinfo {author} {\bibfnamefont {M.~D.}\ \bibnamefont
  {Frye}}\ and\ \bibinfo {author} {\bibfnamefont {J.~M.}\ \bibnamefont
  {Hutson}},\ }\bibfield  {title} {\bibinfo {title} {Characterizing quasibound
  states and scattering resonances},\ }\href
  {https://doi.org/10.1103/PhysRevResearch.2.013291} {\bibfield  {journal}
  {\bibinfo  {journal} {Phys. Rev. Res.}\ }\textbf {\bibinfo {volume} {2}},\
  \bibinfo {pages} {013291} (\bibinfo {year} {2020})}\BibitemShut {NoStop}%
\bibitem [{\citenamefont {Gao}(2000)}]{Gao:2000}%
  \BibitemOpen
  \bibfield  {author} {\bibinfo {author} {\bibfnamefont {B.}~\bibnamefont
  {Gao}},\ }\bibfield  {title} {\bibinfo {title} {Zero-energy bound or
  quasibound states and their implications for diatomic systems with an
  asymptotic van der {W}aals interaction},\ }\href
  {https://doi.org/10.1103/PhysRevA.62.050702} {\bibfield  {journal} {\bibinfo
  {journal} {Phys. Rev. A}\ }\textbf {\bibinfo {volume} {62}},\ \bibinfo
  {pages} {050702(R)} (\bibinfo {year} {2000})}\BibitemShut {NoStop}%
\bibitem [{Note2()}]{Note2}%
  \BibitemOpen
  \bibinfo {note} {Nuclear spin on the Yb atom may cause Feshbach resonances
  similar to those in alkali-metal plus closed-chell atom collisions \cite
  {Brue:LiYb:2012, Barbe:RbSr:2018, Yang:CsYb:2019}, but these are expected to
  be very narrow, so we consider only nuclear spin on the Cr.}\BibitemShut
  {Stop}%
\bibitem [{\citenamefont {Reinhardt}\ \emph {et~al.}(1995)\citenamefont
  {Reinhardt}, \citenamefont {Maichel}, \citenamefont {Baumann},\ and\
  \citenamefont {Kr{\"u}ger}}]{Reinhardt:Cr_HF:1995}%
  \BibitemOpen
  \bibfield  {author} {\bibinfo {author} {\bibfnamefont {T.}~\bibnamefont
  {Reinhardt}}, \bibinfo {author} {\bibfnamefont {J.}~\bibnamefont {Maichel}},
  \bibinfo {author} {\bibfnamefont {M.}~\bibnamefont {Baumann}},\ and\ \bibinfo
  {author} {\bibfnamefont {J.}~\bibnamefont {Kr{\"u}ger}},\ }\bibfield  {title}
  {\bibinfo {title} {Hyperfine structure of the resonance lines of $^{53}${Cr}
  and lifetimes of some excited states of the {Cr I} spectrum},\ }\href
  {https://doi.org/10.1007/BF01439381} {\bibfield  {journal} {\bibinfo
  {journal} {Z. Phys. D}\ }\textbf {\bibinfo {volume} {34}},\ \bibinfo {pages}
  {87} (\bibinfo {year} {1995})}\BibitemShut {NoStop}%
\bibitem [{\citenamefont {Miyazawa}\ \emph {et~al.}(2022)\citenamefont
  {Miyazawa}, \citenamefont {Inoue}, \citenamefont {Matsui}, \citenamefont
  {Nomura},\ and\ \citenamefont {Kozuma}}]{MiyazawaPRL22}%
  \BibitemOpen
  \bibfield  {author} {\bibinfo {author} {\bibfnamefont {Y.}~\bibnamefont
  {Miyazawa}}, \bibinfo {author} {\bibfnamefont {R.}~\bibnamefont {Inoue}},
  \bibinfo {author} {\bibfnamefont {H.}~\bibnamefont {Matsui}}, \bibinfo
  {author} {\bibfnamefont {G.}~\bibnamefont {Nomura}},\ and\ \bibinfo {author}
  {\bibfnamefont {M.}~\bibnamefont {Kozuma}},\ }\bibfield  {title} {\bibinfo
  {title} {Bose-einstein condensation of europium},\ }\href
  {https://doi.org/10.1103/PhysRevLett.129.223401} {\bibfield  {journal}
  {\bibinfo  {journal} {Phys. Rev. Lett.}\ }\textbf {\bibinfo {volume} {129}},\
  \bibinfo {pages} {223401} (\bibinfo {year} {2022})}\BibitemShut {NoStop}%
\end{thebibliography}%


\begin{thebibliography}{5}%
\makeatletter
\providecommand \@ifxundefined [1]{%
 \@ifx{#1\undefined}
}%
\providecommand \@ifnum [1]{%
 \ifnum #1\expandafter \@firstoftwo
 \else \expandafter \@secondoftwo
 \fi
}%
\providecommand \@ifx [1]{%
 \ifx #1\expandafter \@firstoftwo
 \else \expandafter \@secondoftwo
 \fi
}%
\providecommand \natexlab [1]{#1}%
\providecommand \enquote  [1]{``#1''}%
\providecommand \bibnamefont  [1]{#1}%
\providecommand \bibfnamefont [1]{#1}%
\providecommand \citenamefont [1]{#1}%
\providecommand \href@noop [0]{\@secondoftwo}%
\providecommand \href [0]{\begingroup \@sanitize@url \@href}%
\providecommand \@href[1]{\@@startlink{#1}\@@href}%
\providecommand \@@href[1]{\endgroup#1\@@endlink}%
\providecommand \@sanitize@url [0]{\catcode `\\12\catcode `\$12\catcode
  `\&12\catcode `\#12\catcode `\^12\catcode `\_12\catcode `\%12\relax}%
\providecommand \@@startlink[1]{}%
\providecommand \@@endlink[0]{}%
\providecommand \url  [0]{\begingroup\@sanitize@url \@url }%
\providecommand \@url [1]{\endgroup\@href {#1}{\urlprefix }}%
\providecommand \urlprefix  [0]{URL }%
\providecommand \Eprint [0]{\href }%
\providecommand \doibase [0]{https://doi.org/}%
\providecommand \selectlanguage [0]{\@gobble}%
\providecommand \bibinfo  [0]{\@secondoftwo}%
\providecommand \bibfield  [0]{\@secondoftwo}%
\providecommand \translation [1]{[#1]}%
\providecommand \BibitemOpen [0]{}%
\providecommand \bibitemStop [0]{}%
\providecommand \bibitemNoStop [0]{.\EOS\space}%
\providecommand \EOS [0]{\spacefactor3000\relax}%
\providecommand \BibitemShut  [1]{\csname bibitem#1\endcsname}%
\let\auto@bib@innerbib\@empty
%</preamble>
\bibitem [{\citenamefont {{Le Roy}}\ and\ \citenamefont
  {Henderson}(2007)}]{LeRoy:2007}%
  \BibitemOpen
  \bibfield  {author} {\bibinfo {author} {\bibfnamefont {R.~J.}\ \bibnamefont
  {{Le Roy}}}\ and\ \bibinfo {author} {\bibfnamefont {R.~D.~E.}\ \bibnamefont
  {Henderson}},\ }\bibfield  {title} {\bibinfo {title} {A new potential
  function form incorporating extended long-range behaviour: application to
  ground-state {Ca}$_2$},\ }\href {https://doi.org/10.1080/00268970701241656}
  {\bibfield  {journal} {\bibinfo  {journal} {Mol. Phys.}\ }\textbf {\bibinfo
  {volume} {105}},\ \bibinfo {pages} {663} (\bibinfo {year}
  {2007})}\BibitemShut {NoStop}%
\bibitem [{\citenamefont {Le~Roy}\ \emph {et~al.}(2006)\citenamefont {Le~Roy},
  \citenamefont {Huang},\ and\ \citenamefont {Jary}}]{LeRoy:N2:2006}%
  \BibitemOpen
  \bibfield  {author} {\bibinfo {author} {\bibfnamefont {R.~J.}\ \bibnamefont
  {Le~Roy}}, \bibinfo {author} {\bibfnamefont {Y.}~\bibnamefont {Huang}},\ and\
  \bibinfo {author} {\bibfnamefont {C.}~\bibnamefont {Jary}},\ }\bibfield
  {title} {\bibinfo {title} {An accurate analytic potential function for
  ground-state {N}$_2$ from a direct-potential-fit analysis of spectroscopic
  data},\ }\href {https://doi.org/10.1063/1.2354502} {\bibfield  {journal}
  {\bibinfo  {journal} {J. Chem. Phys.}\ }\textbf {\bibinfo {volume} {125}},\
  \bibinfo {pages} {164310} (\bibinfo {year} {2006})}\BibitemShut {NoStop}%
\bibitem [{\citenamefont {Hajigeorgiou}\ and\ \citenamefont
  {Le~Roy}(2000)}]{Hajigeorgiou:2000}%
  \BibitemOpen
  \bibfield  {author} {\bibinfo {author} {\bibfnamefont {P.~G.}\ \bibnamefont
  {Hajigeorgiou}}\ and\ \bibinfo {author} {\bibfnamefont {R.~J.}\ \bibnamefont
  {Le~Roy}},\ }\bibfield  {title} {\bibinfo {title} {{A ``modified
  Lennard-Jones oscillator'' model for diatom potential functions}},\ }\href
  {https://doi.org/10.1063/1.480946} {\bibfield  {journal} {\bibinfo  {journal}
  {J. Chem. Phys.}\ }\textbf {\bibinfo {volume} {112}},\ \bibinfo {pages}
  {3949} (\bibinfo {year} {2000})}\BibitemShut {NoStop}%
\bibitem [{\citenamefont {Tomza}(2013)}]{Tomza:2013}%
  \BibitemOpen
  \bibfield  {author} {\bibinfo {author} {\bibfnamefont {M.}~\bibnamefont
  {Tomza}},\ }\bibfield  {title} {\bibinfo {title} {Prospects for ultracold
  polar and magnetic chromium--closed-shell-atom molecules},\ }\href
  {https://doi.org/10.1103/PhysRevA.88.012519} {\bibfield  {journal} {\bibinfo
  {journal} {Phys. Rev. A}\ }\textbf {\bibinfo {volume} {88}},\ \bibinfo
  {pages} {012519} (\bibinfo {year} {2013})}\BibitemShut {NoStop}%
\bibitem [{\citenamefont {Hutson}\ and\ \citenamefont
  {Le~Sueur}(2022)}]{mbf-github:2022}%
  \BibitemOpen
  \bibfield  {author} {\bibinfo {author} {\bibfnamefont {J.~M.}\ \bibnamefont
  {Hutson}}\ and\ \bibinfo {author} {\bibfnamefont {C.~R.}\ \bibnamefont
  {Le~Sueur}},\ }\href@noop {} {\bibinfo {title} {{\sc molscat}, {\sc bound}
  and {\sc field}, version 2022.0}},\ \bibinfo {howpublished}
  {\url{https://github.com/molscat/molscat}} (\bibinfo {year}
  {2022})\BibitemShut {NoStop}%
\end{thebibliography}%

\end{document}

\draft
\graphicspath{{./Figures/}}

\title{Supplemental Material for: \\
Spin-spin interaction and magnetic Feshbach resonances in collisions of high-spin atoms with closed-shell atoms}

\author{Matthew D. Frye}
\email{matthew.frye@fuw.edu.pl} 
\affiliation{Faculty of Physics, University of Warsaw, Pasteura 5, 02-093 Warsaw, Poland}
\author{Piotr S. {\.Z}uchowski}
\affiliation{Institute of Physics, Faculty of Physics, Astronomy and Informatics, Nicolaus Copernicus University, ul.\ Grudziadzka 5/7, 87-100
Torun, Poland}
\author{Micha{\l} Tomza}
\email{michal.tomza@fuw.edu.pl} 
\affiliation{Faculty of Physics, University of Warsaw, Pasteura 5, 02-093 Warsaw, Poland}

\date{\today}

\maketitle

\section{Fitting the interaction potential and coupling}

We represent the interaction potential using a Morse-long-range form, as described by Le Roy and Henderson \cite{LeRoy:2007}. This is a generalization of the Morse potential to include a long-range potential of specified form, $u_\textrm{LR}(R)$, which is usually inverse power. The potential is given by
\begin{equation}
V_\textrm{MLR}(R)=D_\textrm{e}\left(\left\{ 1- \frac{u_\textrm{LR}(R)}{u_\textrm{LR}(R_\textrm{e})} \exp\left[-\phi(R)y_p(R)\right] \right\}^2-1\right),
\end{equation}
where $D_\textrm{e}$ is the depth of the potential at $R_\textrm{e}$, the position of the minimum. $y_p(R)$ is an auxiliary radial variable
\begin{equation}
y_p(R)=\frac{R^p-R_\textrm{e}^p}{R^p+R_\textrm{e}^p},
\end{equation}
and $\phi(R)$ is expanded as a polynomial in $y_p(R)$. Following Refs.\ \cite{LeRoy:N2:2006, LeRoy:2007}, it is convenient to choose a form with $\phi_\infty=\lim_{R\to\infty}\phi(R)$ as a parameter,
\begin{equation}
\phi(R)=\left[1-y_p(R)\right]\sum_{i=0}^N\phi_iy_p(R)^i+y_p(R)\phi_\infty.
\end{equation}
In this way we can choose $\phi_\infty=\ln\left[2D_\textrm{e}/u_\textrm{LR}(R_\textrm{e})\right]$, to ensure the potential correctly approaches $u_\textrm{LR}(R)$ as $R\to\infty$.

We include only one leading term in the long-range potential $u_\textrm{LR}(R)=C_6R^{-6}$. This form then corresponds to the ``modified Lennard-Jones'' of Ref.\ \cite{Hajigeorgiou:2000}, but it is more convenient to use the more general form of Ref.\ \cite{LeRoy:2007}.
We choose $p=2$ to ensure the next leading term behaves as $R^{-8}$ asymptotically, although we do not control its coefficient. We pick $N=2$, which we find gives a satisfactory representation of the shape of the potential.

\begin{table}[htbp]
\caption{Interaction potential for Cr--Yb from Ref.\ \cite{Tomza:2013}. \label{table:points}} \centering
\begin{ruledtabular}
\begin{tabular}{ccc}
i	& $R_i$ ($a_0$)	& $V_i/hc$ cm$^{-1}$\\
\hline
1 & 4.2 &   10653 \\
2 & 4.4 &    6807.0 \\
3 & 4.6 &  3866.8 \\
4 & 4.8 &    1654.0 \\
5 & 5.0 &  27.241 \\
6 & 5.2 & $-1131.3$ \\
7 & 5.4 & $-2039.9$ \\
8 & 5.6 & $-2513.5$ \\
9 & 5.8 & $-2768.7$ \\
10 & 6.0 & $-2862.3$ \\
11 & 6.2 & $-2840.1$ \\
12 & 6.4 & $-2737.9$ \\
13 & 6.6 & $-2584.0$ \\
14 & 6.8 & $-2399.8$ \\
15 & 7.0 & $-2201.4$ \\
16 & 7.2 & $-2000.0$ \\
17 & 7.4 & $-1803.8$ \\
18 & 7.6 & $-1617.8$ \\
19 & 8.0 & $-1287.1$ \\
20 & 8.5 & $-956.56$ \\
21 & 9.0 & $-708.75$ \\
22 & 9.5 & $-526.37$ \\
23 & 10.0 & $-393.07$ \\
24 & 11.0 & $-223.57$ \\
25 & 12.0 & $-130.55$ \\
26 & 13.0 & $-78.298$ \\
27 & 14.0 & $-48.265$ \\
28 & 15.0 & $-30.591$ \\
29 & 16.0 & $-19.927$ \\
30 & 17.0 & $-13.338$ \\
31 & 18.0 & $-9.1525$ \\
32 & 19.0 & $-6.4305$ \\
33 & 20.0 & $-4.6147$ \\
34 & 22.0 & $-2.5063$ \\
35 & 24.0 & $ -1.4450$ \\
36 & 26.0 & $-0.87292$ \\
37 & 28.0 & $-0.55213$ \\
38 & 30.0 & $-0.35728$ \\
39 & 35.0 & $-0.13498$ \\
40 & 40.0 & $-0.058578$ \\
41 & 50.0 & $-0.012203$
\end{tabular}
\end{ruledtabular}
\end{table}

We use a standard least-squares fitting procedure to fit $V_\textrm{MLR}(R)$ to \emph{ab initio} calculations for the Cr--Yb potential from Ref.\ \cite{Tomza:2013}. The full potential was not published with that paper, so we tabulate the values for reference in Table \ref{table:points}. In the absence of meaningful uncertainties from the \emph{ab initio} calculations, we assign
\begin{equation}
\sigma_i=\begin{cases}
|V_i/D_\textrm{e}|, 		& R_i>R_\textrm{e} \\
|V_i/D_\textrm{e}+2|, 	& R_i<R_\textrm{e}
\end{cases},
\end{equation}
such that outside $R_\textrm{e}$ the calculations are assumed to have a constant fractional uncertainty, and inside $R_\textrm{e}$ the weights decrease monotonically.

\begin{figure}[tbp]
\includegraphics[width=0.99\linewidth]{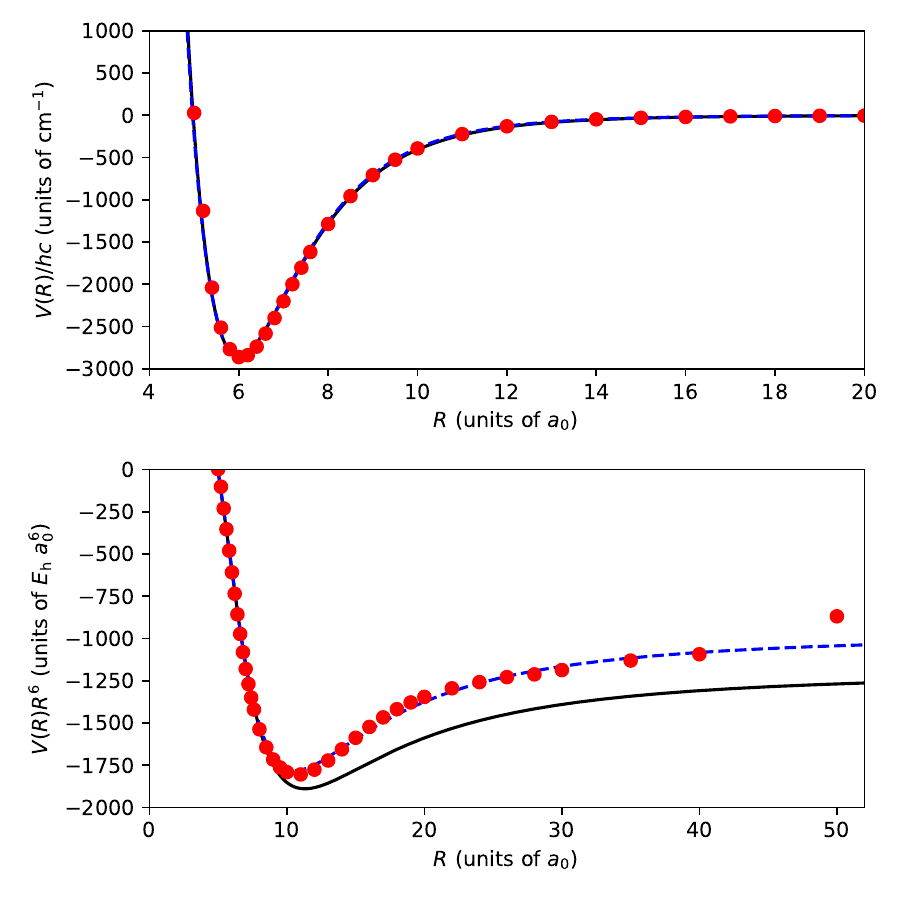}
\caption{Electronic interaction potential $V_\textrm{elec}(R)$ for Cr+Yb. Red dots show ab initio calculations of Ref.\ \cite{Tomza:2013}, dashed blue lines show the initial fit with $C_6$ allowed to vary, and solid black lines show the final potential with the accurate $C_6$ reimposed. Potential is shown as both $V_\textrm{elec}(R)$ (top) and $V_\textrm{elec}(R)\times R^6$ (bottom), which emphasizes the important long-range part of the potential.
\label{fig:potential}}
\end{figure}

In the fitting we fix $R_\textrm{e}=R_{10}=6$ $a_0$, and $D_\textrm{e}=|V_{10}|=2862.3 \times hc$ cm$^{-1}$. We have a good \emph{ab initio} estimate of $C_6=1195$ a.u.\ \cite{Tomza:2013}, but we allow $C_6$ to vary in the fit as the \emph{ab initio} points are better described by a $C_6$ about 20\% smaller. We also find that the outermost point $R_{41}=50$ $a_0$ does not follow the potential implied by the other points, so we exclude it from our fit. 
The fitted parameters are $(\phi_0,\phi_1,\phi_2)=(-1.873,0.333,-0.621)$ and $C_6=971$ a.u. We obtain our final potential by reimposing $C_6=1195$ a.u. These results are shown in Fig.\ \ref{fig:potential}, both as the potential itself, $V(R)$, and as $V(R)\times R^6$, which is better for visualizing the long-range behavior.

The spin--spin coupling is calculated as described in the main text, and the results are tabulated in Table \ref{table:lambda_points}. We fit these to a double exponential decay $\lambda(R)=A e^{-\beta R}+B e^{-2\beta R}$, and find $\beta=0.48$ $a_0^{-1}$, $A=0.056$ cm$^{-1}$, and $B=50$ cm$^{-1}$. This is shown in Fig.\ 1 of the main paper.

\begin{table}[htbp]
\caption{Values of $\lambda(R)$, calculated as described in the main text. \label{table:lambda_points}} \centering
\begin{ruledtabular}
\begin{tabular}{ccc}
i	& $R$ (\AA)	& $\lambda(R)/hc$ cm$^{-1}$\\
\hline
	1	& 3.00	& 0.357743	\\
	2	& 3.25	& 0.262367	\\
	3	& 3.50	& 0.187532	\\
	4	& 3.75	& 0.129186	\\
	5	& 4.00	& 0.086068	\\
	6	& 4.25	& 0.056202	\\
	7	& 4.50	& 0.036166	\\
	8	& 4.75	& 0.023105	\\
	9	& 5.00	& 0.014657	\\
	10	& 5.25	& 0.009296	\\
	11	& 5.50	& 0.005909	\\
	12	& 5.75	& 0.003773	\\
	13	& 6.00	& 0.002396	\\
	14	& 6.25	& 0.001527	\\
	15	& 6.50	& 0.000992	\\
	16	& 7.00	& 0.000459	\\
	17	& 7.50	& 0.000235	\\
	18	& 8.00	& 0.000140	\\
	19	& 8.50	& 0.000080	\\
	20	& 9.00	& 0.000045	\\
	21	& 9.50	& 0.000026	\\
	22	& 10.00	& 0.000015
\end{tabular}
\end{ruledtabular}
\end{table}

\section{Example field scans}

\begin{figure}[tbp]
\includegraphics[width=0.99\linewidth]{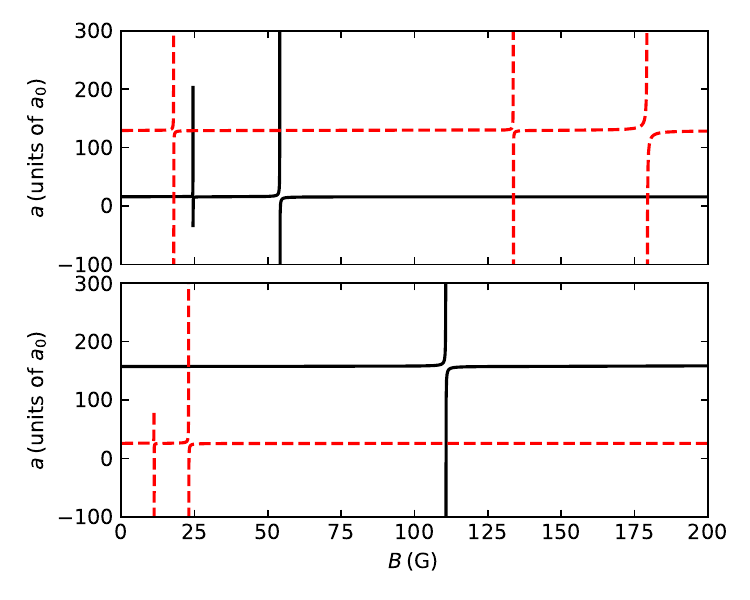}
\caption{Example magnetic field scans of scattering length, showing typical resonances. Black solid lines and red dashed lines show potential number 15 and 65, respectively. Top panel shows $^{52}\textrm{Cr}+{^{174}\textrm{Yb}}$ and bottom panel shows $^{53}\textrm{Cr}+{^{174}\textrm{Yb}}$.
\label{fig:scans}}
\end{figure}

Figure \ref{fig:scans} shows example scans of scattering length as a function of magnetic field up to 200 G, showing typical resonances. We use two potentials: number 15 with $a\approx 2.9 \bar{a}$ for $^{52}\textrm{Cr}+{^{174}\textrm{Yb}}$, and number 65 with $a\approx 0.5 \bar{a}$ for $^{52}\textrm{Cr}+{^{174}\textrm{Yb}}$. We also show results for $^{53}\textrm{Cr}+{^{174}\textrm{Yb}}$ for the same potentials.

\section{Additional files}

\subsection{Details of coupled-channels calculations}

As stated in the main text, we perform coupled-channels calculations using \textsc{molscat}, \textsc{bound}, and \textsc{field}. We provide a number of additional files relating to these calculations attached to this Supplemental Material, all of which should be read with reference to the user manual \cite{mbf-github:2022}. The files \path{base9-Sat_1S.f} and \path{vstar-Sat_1S.f} are plug-in routines for basis-set construction and interaction potential input, respectively. \path{mol_CrYb.inp} and \path{fld_CrYb.inp} are sample input files for \textsc{molscat} and \textsc{field}, respectively.
%We also used a script developed previously in Ref.\ \cite{Frye:DyYb:2020} to automate the necessary calculations for locating and characterising the large number of resonances.

\subsection{Tabulated resonance parameters}

The files \path{52_174_resonances.txt}, \path{52_170_resonances.txt}, \path{53_174_resonances.txt}, and \path{53_170_resonances.txt} contain resonance parameters for respectively $^{52}\textrm{Cr}+{^{174}\textrm{Yb}}$, $^{52}\textrm{Cr}+{^{170}\textrm{Yb}}$, $^{53}\textrm{Cr}+{^{174}\textrm{Yb}}$, and $^{53}\textrm{Cr}+{^{170}\textrm{Yb}}$. Tabulations give an index of the potential, the potential depth $D_\textrm{e}$, an index of the bound state causing the resonance, the resonance position $B_\textrm{res}$, the resonance width $\Delta$, the background scattering length $a_\textrm{bg}$, and the normalized width $\bar{\Delta}$.

\bibliography{../CrYb_resonance}